               \documentclass[preprint,review,12pt, authoryear]{elsarticle}

\usepackage{times}
\usepackage{graphicx}
\usepackage{caption}
\usepackage{subcaption}

\usepackage{amssymb}
\usepackage{amsmath}
\usepackage[title,titletoc,toc]{appendix}

\usepackage[nodots,nocompress]{numcompress}

\usepackage{lineno}
\usepackage{ctable}
\usepackage{rotating}
\usepackage[pdftex,colorlinks=false,bookmarks,citecolor=black,%
  bookmarksnumbered,bookmarksopen=true,breaklinks,%
  hyperfootnotes=false,linkcolor=blue]{hyperref}
\newif\ifincludegraphics
\includegraphicstrue 

\usepackage{epstopdf}
\biboptions{semicolon, square,comma,numbers,super,sort&compress}

\journal{icarus}

\begin{document}

\begin{frontmatter}



\title{Water Ice and Dust in the Innermost Coma of Comet 103P/Hartley 2}


\author[label1]{Silvia Protopapa}
\author[label1]{Jessica M. Sunshine}
\author[label1]{Lori M. Feaga}
\author[label1]{Michael  S. P. Kelley}
\author[label1]{Michael  F. A' Hearn}
\author[label1]{Tony L. Farnham}
\author[label2]{Olivier Groussin}
\author[label1,label3]{Sebastien Besse}
\author[label4]{Fr\'{e}d\'{e}ric Merlin}
\author[label5]{Jian-Yang  Li}

\address[label1]{Department of Astronomy, University of Maryland, College Park, MD, 20742, USA}
\address[label2]{Aix-Marseille Universit\'{e}, CNRS, LAM (Laboratoire dÕAstrophysique de Marseille) UMR 7326, 13388, Marseille, France}
\address[label3]{ESA/ESTEC, Keplerlaan 1, Noordwijk, The Netherlands}
\address[label4]{Universit\'{e} Denis Diderot Paris 7, LESIA, Observatoire de Paris, France}
\address[label5]{Planetary Science Institute, Tucson, AZ 85719, USA}
\begin{abstract}
On November 4th, 2010, the Deep Impact eXtended Investigation (DIXI) successfully encountered comet 103P/Hartley 2, when it was at a heliocentric distance of 1.06 AU. Spatially resolved near-IR spectra of comet Hartley 2 were acquired in the 1.05 -- 4.83 $\mu$m wavelength range using the HRI-IR spectrometer. We present spectral maps of the inner $\sim$10 kilometers of the coma collected 7 minutes and 23 minutes after closest approach. The extracted reflectance spectra include well-defined absorption bands near 1.5, 2.0, and 3.0 $\mu$m consistent in position, bandwidth, and shape with the presence of water ice grains. Using Hapke's radiative transfer model, we characterize the type of mixing (areal vs. intimate), relative abundance, grain size, and spatial distribution of water ice and refractories. Our modeling suggests that the dust, which dominates the innermost coma of Hartley 2 and is at a temperature of 300K, is thermally and physically decoupled from the fine-grained water ice particles, which are on the order of 1 $\mu$m in size. The strong correlation between the water ice, dust, and CO$_{2}$ spatial distribution supports the concept that CO$_{2}$  gas drags the water ice and dust grains from the nucleus. Once in the coma, the water ice begins subliming while the dust is in a constant outflow. The derived water ice scale-length is compatible with the lifetimes expected for 1-$\mu$m pure water ice grains at 1 AU, if velocities are near 0.5 m/s. Such velocities, about three order of magnitudes lower than the expansion velocities expected for isolated 1-$\mu$m water ice particles \citep{Hanner1981,Whipple2}, suggest that the observed water ice grains are likely aggregates.
\end{abstract}


\end{frontmatter}
\section{Introduction}
Comets formed beyond the H$_{2}$O frost line, where ices can condense \citep[\textit{e.g.}, ][]{A'Hearn2012ApJ}. A variety of processes have affected comets during their long storage in the Oort cloud and scattered disk (\textit{e.g.}, irradiation by energetic particles, heating by passing stars, collisions). Similarly, repeated solar heating leads to evolution of short period comets during their many passages close to the Sun.  However, because most of these processes affect only the outer layer of comets, the pristine nature of the bulk of the nucleus is preserved \citep{Stern2003, Mumma1993,Weissman1997,Morvan2004}. As such, comets are excellent laboratories to extend our understanding of the origin and evolution of
the Solar System. 

Given that comets contain the least processed primordial materials that formed the cores of the giant planets \citep{AHearn2011a}, the analysis of the composition and physical state of cometary materials
is critical to improve our understanding of the accretion processes that led to the formation of comet nuclei and ultimately the planets. Water is a key component of comets \citep{Mumma_review2011, Feaga2007, Morvan2004, Morvan1997} and water ice has been observed from the ground and with
\textit{in situ} observations  in the comae  \citep{Davies1997, Kawakita2004,Yang2009, Sunshine2011a}, on the surfaces \citep{Sunshine2006}, and in the near-surface interiors \citep{Sunshine2007} of comets. An example of addressing comet nuclei formation by means of water ice characteristics is given by \citet{Sunshine2007}: the presence of very fine ($\sim$1 $\mu$m) water ice particles in the impact ejecta of comet Tempel 1, free of refractory impurities, led to question the interstellar dust grain model proposed by \citet{Greenberg1998} and \citet{Greenberg1999}. The basic idea of this model is that comets formed directly through coagulation of interstellar dust. As such, the morphological structure of comet nuclei is an aggregate of presolar interstellar dust grains, which consist of a core of silicates mantled first by a shell of organic refractory material and then by a mixture of water dominated ices, which are embedded with thousands of very small (1-10 nm) carbonaceous/large molecule particles. According to \citet{Sunshine2007},  it is unlikely that water ice would segregate from the superfine particles in the impact ejecta. The model by Greenberg is only one of the several models proposed, describing  the underlying structure of cometary nuclei (\textit{e.g.}, the ``fluffy aggregate'' of \citet{Donn1985} and \citet{Donn1986}, the ``rubble pile model'', either collisionally modified \citep{Weissman2004} or primordial  \citep{Weissman1986}, the ``icy-glue'' model of \citet{Gombosi1986}, the ``talps'' or ``layered pile'' model by \citet{Belton2007}). All of these models consider different origins and/or evolutionary processes. Investigations of cometary nuclei at close range, which can only be achieved by space missions, are required to validate, disprove, or improve these various formation models.

On November 4th, 2010, the Deep Impact Flyby (DIF) spacecraft \citep{A'Hearn2005,Hampton2005} successfully encountered comet 103P/Hartley 2 as part of its extended mission, the Deep Impact eXtended Investigation (DIXI) \citep{Hearn2011}. The DIF spacecraft employs two multi-spectral imagers (MRI-VIS, HRI-VIS) and a 1.05 -- 4.83 $\mu$m  near-infrared spectrometer (HRI-IR). Near-IR spectra of comet Hartley 2 were acquired for several weeks before and after closest approach (CA, at a range of 694 km, occurred on November 4th, 2010, at 13:59:47.31 UTC with a maximum HRI-IR spatial resolution of 7 m/pixel). DIXI observations revealed a bi-lobed, very small (maximum length of 2.33 km, \citep{Hearn2011}), and highly active nucleus, with water ice distributed heterogeneously, in specific areas on the surface \citep{Sunshine2011c} and in the coma \citep{Sunshine2011a,Protopapa2011}. 

In this paper, we present a detailed analysis of the composition and texture of water ice and refractories in the inner-most coma of Hartley 2, within a few kilometers of the surface, using DI HRI-IR data. We investigate the physical makeup of the water ice grains, with the goal of providing more observational constraints to help us have a better understanding of the accretion process that led to the formation of comet nuclei. We investigate the role of the gas in delivering the water ice and dust to the coma of Hartley 2 and how the emitted material evolves after it leaves the nucleus.
\section{Observations}\label{Observations and data reduction}
Spatially resolved near-IR spectra of comet Hartley 2 were acquired in the 1.05 -- 4.83 $\mu$m wavelength range using the HRI-IR spectrometer \citep{Hampton2005}. HRI-IR data were collected from 01 October 2010 to 26 November 2010. In this paper, only a subset of data is analyzed. In particular, we present HRI-IR spatially resolved scans of comet Hartley 2 collected 7 (ID 5006000) and 23 (ID 5007002) minutes post-CA. Table \ref{tabA} lists the main characteristics of these two scans.

The IR data are calibrated using the DI science data pipeline \citep{Klaasen2008,Klaasen2012} and are available in the NASA Planetary Data System (PDS) archive \citep{McLaughlin2013}. The standard steps of the pipeline processing include linearization, dark subtraction, flat fielding, conversion from DN to radiance units (W m$^{-2}$sr$^{-1}$$\mu$m$^{-1}$), and bad pixel masking. These data have been recalibrated since the work by \citet{Hearn2011}, although the differences are not dramatic. The 5006000 and 5007002 radiance maps are shown in panels (a) and (b) of Figure \ref{radiance-maps}, respectively, compared with the HRIVIS and MRI-VIS context images (with characteristics
given in Table \ref{tab_context_images}). Because the HRI-IR instrument is a scanning slit spectrometer, reconstructed spatial information is susceptible to errors from jumps in spacecraft attitude due to impacting particles or autonomous
operations.  A mismatch in the nucleus shape between the IR scan acquired 7 min post-CA and the corresponding visible context images was observed, possibly caused by very small grains hitting the spacecraft at high speed. We have corrected this mismatch by shifting the bottom 14 rows of the 56-row IR scan by 1 pixel to the left and the top 21 rows by 1 pixel toward the right and bottom (with respect to the orientation of Figure \ref{radiance-maps}). 

Both the 5006000 and 5007002 scans contain the nucleus of Hartley 2 and the surrounding coma. The nucleus is in roughly the same orientation in the two scans. The precession of the long axis of the nucleus around the angular momentum vector has a period of 18.4 hours at encounter \citep{Hearn2011,Belton2013}; therefore in 16 min separating the two scans, the rotation of the nucleus is negligible with regards to coma studies, as is the change of the spacecraft line of sight.  In both scans, there are jets off the end of the smaller lobe of the nucleus and beyond the terminator along the lower edge of the larger lobe (Figure \ref{radiance-maps}).

The nucleus of Hartley 2 needs to be masked in order to focus our analysis on the water ice and refractories in the inner-most coma.  The dashed red line in Figure \ref{radiance-maps} outlines the illuminated portion of the cometary nucleus, including the scattered light near the limb, that is excluded from our
analysis. This mask was defined at long wavelengths ($\lambda$ $\geq$ 3.5 $\mu$m), where the thermal
emission of the nucleus dominates over the signal from the dust in the
coma, allowing the nucleus and coma to be disentangled.
The nucleus contour is defined by radiance values in the 450 channel ($\sim$4.1 $\mu$m) greater than 0.16 W m$^{-2}$sr$^{-1}$$\mu$m$^{-1}$ in order to match the nucleus as seen in the HRI-VIS and MRI-VIS images acquired at the same time as the IR data. 
\section{Spectral Modeling}
\subsection{Reflectance}\label{From radiance to reflectance}

Once the flux is calibrated into radiance, $I$, from the pipeline, we can convert it into reflectance, $R$, by
\begin{equation}
R(\lambda) = \frac{\pi I(\lambda) r^{2}}{F_{\odot}(\lambda)}
\end{equation}
where $F_{\odot}(\lambda)$ is the solar flux at 1 AU, and $r$ is the heliocentric distance, in our case 1.06 AU. The solar spectrum we use is a synthesized spectrum from various sources \citep{Berk2006,Kurucz1995}, available at the MODTRAN web site \url{http://rredc.nrel.gov/solar/spectra/am0/other_spectra.html}. 

The next step of our analysis is the determination of the error on the radiance. The derived error will be used in the best fit minimization of the reflectance spectra (see Section \ref{Results}). At each wavelength, the radiance map (see Figure \ref{errorRadiance}, panel a) is smoothed using a moving resistant mean in a  3$\times$3 pixel box (see Figure \ref{errorRadiance}, panel b). Outliers that  vary by more than 2.5$\sigma$ are excluded. The smoothed radiance map is then subtracted from the original one, producing a background noise map (see Figure \ref{errorRadiance}, panel c) where coma structures are no longer visible, except where the jets are particularly strong (\textit{e.g.}, jets originating from the lower edge of the larger lobe of the nucleus).  For each pixel of the background noise map, the standard deviation of the mean of its neighboring pixels located in a 3-by-3 pixel box is computed and assumed to be the radiance error. A careful examination of panel (c) in Figure \ref{errorRadiance} allows us to conclude that this approach can be applied everywhere except for areas with coma structures that still remain in the background map. Avoiding these areas we take the mean uncertainty of the background at each wavelength (averaged over all pixels within the white boxes in panel c of Figure \ref{errorRadiance}) as the uncertainty for all pixels in the image. This approach computes the intrinsic radiance error, dominated by warm bench background \citep[for a detailed description of the sources of error in the HRI-IR instrument, the reader is referred to][]{Klaasen2008,Klaasen2012}. The error ranges from a few percent at short wavelengths to a fraction of a percent at long wavelengths. The uncertainty in the absolute instrument calibration is 10\% \citep{Klaasen2012}. However, the spectral fitting analysis is performed without considering this 10\% absolute error.

\subsection{Dominant Spectral Features}\label{Spectral Features}
Figure \ref{continuum_modeling} shows the comparison between two reflectance spectra extracted inside and outside the jet emerging from the small lobe of the nucleus facing the Sun as observed 7 min post-CA. Both spectra in Figure \ref{continuum_modeling} are red sloped (increasing reflectance with increasing wavelength) from 1.2 -- 2.5 $\mu$m. Specifically, the spectra extracted in boxes A (top panel of Figure \ref{continuum_modeling}) and B (bottom panel of Figure \ref{continuum_modeling}) of Figure \ref{errorRadiance} (panel a) have a spectral slope of 5.10$\pm$0.08\%/0.1$\mu$m  and 1.41$\pm$0.02\%/0.1$\mu$m, respectively.  In general, red slopes are well reproduced by refractory components (\textit{e.g.}, amorphous carbon) \citep{Campins2006}. Beyond 2.5 $\mu$m, the spectrum is the sum of two components, one from scattered sunlight and one from thermal emission. Several emission features can also be seen in the reflectance spectra. These include emission from H$_{2}$O vapor at 2.7 $\mu$m, CO$_{2}$ at 4.3 $\mu$m, and organics in the 3.3 -- 3.5 $\mu$m region. These emission features will not be discussed further in this paper, as they are addressed by \citet{Feaga2012} and \citet{Besse2012}. While most areas in the coma do not show significant absorption features (Figure \ref{continuum_modeling}, top panel),  the spectra extracted in the jet region (Figure \ref{continuum_modeling}, bottom panel) clearly display water ice absorption bands at 1.5, 2.0, and 3.0 $\mu$m. Water ice is blue in the near infrared, which explains  the lower spectral slope of the water ice-rich spectrum with respect to the water ice-depleted one. The region sampled to display the water ice-rich spectrum in this paper (box B) is similar to that discussed by \citet{Hearn2011} in his Figure 5 (bottom right panel). Unlike box B, box A is displaced from the region sampled by \citet{Hearn2011} to display the water ice-depleted spectrum (see Section \ref{Results}).

\subsection{Modeling Analysis}\label{Modeling}
The flux from the coma in the wavelength range covered by the HRI-IR spectrometer is the sum of the scattered ($F_{scat}$) and thermal ($F_{thermal}$) components, the latter being dominant beyond 3.4 $\mu$m (see dash-dot orange line in Figure \ref{continuum_modeling}). 
\subsubsection{Scattered Component}
The scattered flux is calculated at each wavelength via 
\begin{equation}
F_{scat}(\lambda) = \frac{F_{\odot}(\lambda)p(\lambda)\Phi(\alpha)C}{\pi r^{2}\Delta^{2}} 
\end{equation}
where $p(\lambda)$ is the geometric albedo of each single grain in the coma, $\Phi(\alpha)$ is the scattering phase function, $\Delta$[m] is the spacecraft-comet distance, and the scale factor $C$[m$^{2}$] represents the total geometric cross section of the cometary grains within the slit. 
The contribution of the scattered flux to the reflectance, indicated as $R_{scat}$, is given by
\begin{equation}
R_{scat}(\lambda)=f\Phi(\alpha)p(\lambda)
\end{equation}
where $f$ is the filling factor of the grains in the field of view (see \ref{scattered flux}).
The coma is a collection of many individual grains. In this analysis all grains are considered to be identical, each with geometric albedo $p(\lambda)$, computed by means of the \citet{Hapke1993} radiative transfer model. We adopt the standard approach that a single grain is composed of independent scatterers, or particles.  The qualitative spectroscopic analysis presented in Section \ref{Spectral Features} suggests the presence of two particle types: water ice and refractories. Two modes of mixing are possible: areal and intimate. Areal mixing (also called linear mixing) assumes that the different types of particles are spatially well separated and that a photon does not encounter different particle composition in one scattering event. In an intimate mixture, the particles are mixed homogeneously together in close proximity allowing for multiple scattering. 

In the case of areal mixtures, the geometric albedo of each average grain consists of a linear combination of the geometric albedo of each component $p_{j} $ ($j = $ H$_{2}$O-ice, refractories) weighted by its spatial extent:
\begin{equation}
p(\lambda) = \sum_{j}F_{j}p_{j} (\lambda)
\end{equation}
where  $F_{j}$ is the fraction of the area occupied by the $j$th component, such that $\sum_{j}F_{j}=1$. 
In the case of isotropic scattering we have
\begin{equation}
p_{j}(\lambda) =  \frac{1}{2}r_{j}(\lambda)+\frac{1}{6}r_{j}^{2}(\lambda)
\end{equation}
where $r_{j}(\lambda)$ is the diffusive reflectance of the $j$th component, which is a function of the particle single scattering albedo $w_{j}(\lambda)$ via
\begin{equation}
r_{j}(\lambda) = \frac{1-\sqrt{1-w_{j}(n_{j}(\lambda),k_{j}(\lambda),<D_{j}>)}}{1+\sqrt{1-w_{j}(n_{j}(\lambda),k_{j}(\lambda),<D_{j}>)}}.
\end{equation}
The particle single scattering albedo describes the scattering properties of the material. It is computed by means of the \citet{Hapke1993} equivalent-slab and exponential models for large ($X=\pi D/\lambda >> 1$, where $\lambda$ is the wavelength of the incident radiation and $D$ is the particle diameter) irregular particles. The particle single scattering albedo is a function of the optical constants $n_{j}$ and $k_{j}$, the real and imaginary part of the refractive index, respectively, and $<D_{j}>$, which represents the mean ray path length within the $j$th-type of particle. In the case of spherical particles,  $<D_{j}> \simeq 0.9 D_{j}$. However, if the particles are irregular with maximum dimension $D_{j}$, then $<D_{j}>$ can be quite different than $D_{j}$, and in general will be smaller (\textit{i.e.}, they have smaller mean cross sections than the circumscribed sphere). We will from now on refer to $<D_{j}>$ as particle diameter.

In the case of intimate mixture, the averaging process is on the level of the scattering properties of the various types of particles in the mixture \citep{Hapke1993}. The parameters that enter in the reflectance equation are averages of those of each particle type weighted by cross-sectional area (given by the fractional volume $V$ divided by the size of the grains $D$). We have the following equation for the geometric albedo:
\begin{equation}
p(\lambda) =  \frac{1}{2}r(\lambda)+\frac{1}{6}r^{2}(\lambda),
\end{equation}
where
\begin{equation}
r(\lambda) = \frac{1-\sqrt{1-\overline{w}(\lambda)}}{1+\sqrt{1-\overline{w}(\lambda)}}
\end{equation}
with
\begin{equation}
\overline{w}(\lambda) =\Bigg [\sum_{j}\frac{ V_{j} }{ D_{j} }w_{j} \Bigg ]\Bigg [\sum_{j}\frac{ V_{j} }{ D_{j} } \Bigg ]^{-1}.
\end{equation}
\subsubsection{Thermal Component}
We model the thermal flux density $F_{thermal}$ as
\begin{equation}
F_{thermal}(\lambda) = \frac{\epsilon(\lambda)B(T_{c},\lambda)C_{e}}{\Delta^{2}}
\end{equation}
where $\epsilon(\lambda)$ is the emissivity, $C_{e}$ is the emitting cross section, $T_{c}$[K] is the coma temperature,  and $B$ is the Planck function defined by the following equation
\begin{equation}
B(T_{c},\lambda) = \frac{2hc^{2}}{\lambda^{5}}(e^{\frac{hc}{\lambda k_{b}T_{c}}}-1)^{-1}
\end{equation}  
where $k_{b}$ is  the Boltzmann constant, $h$ the Planck constant, and $c$ the speed of light. The contribution of the thermal flux to the reflectance, indicated as $R_{thermal}$, is given by
\begin{equation}
R_{thermal}(\lambda)=\frac{\pi r^{2}\epsilon(\lambda)B(T_{c},\lambda)f_{e}}{F_{\odot}}
\end{equation}
where $f_{e}$ is the filling factor of the grains.
\subsection{Modeling Results}\label{Results}
In our analysis, to model the observed reflectance spectra, we first consider an areal mixture (the intimate mixing mode is analyzed in Section \ref{Areal vs Intimate}) of crystalline water ice ($T$ = 266 K, optical constants from \citet{Warren2008}) with a dark and featureless refractory component (\textit{e.g.}, amorphous carbon, optical constants from \citet{Edoh1983}). Because there are no absorption features detected in the near-infrared range
other than those of water ice, the non-icy component cannot be unambiguously
identified, and we adopt a material that reproduces the reddening
characteristics.  The variety of compounds that can be used is limited by the
small number of materials whose optical constants have been determined over
the range of wavelengths studied here, but silicates, organics (\textit{e.g.},
tholins), and amorphous carbon are possible candidates \citep{Davies1997,Kawakita2004,Yang2009}.  Amorphous carbon (AC) was
selected because it is featureless and is the lowest albedo reddening
agent. Its characteristics, compared to the other materials, imply that our
results will provide an upper limit on the column density of refractories.

We assume $\epsilon(\lambda)=1$ and neglect $\Phi(\alpha)$, given the small variations of phase angle ranging between 90$^{\circ}$ and 94$^{\circ}$ for data acquired within $\sim$10 hr post-CA. Also, due to the heliocentric distance of comet Hartley 2 (1.06 AU) and the small field of view ($\sim$500$\times$500 $\mu$rad), the phase angle can be considered constant for all grains along the line of sight of the spacecraft. Furthermore, the proper treatment would consist of considering a dust phase function in water ice-depleted regions and a phase
function for aggregates of water ice grains, or comae composed of
water ice and dust in water ice-enriched regions. However, while the former is approximated by \citet{schleicher11}, the latter has not been estimated yet. Thus, we prefer to leave
the phase function as 1.0 for both components, rather than make an
assumption on one or both of these populations.
Therefore, the free parameters in our model are the water ice-to-dust fraction ($F_{H_{2}O}$), the coma temperature $T_{c}$, and the scattering ($f$) and emitting ($f_{e}$) filling factors. They are iteratively modified by means of a $\chi^{2}$ minimization algorithm (Levenberg-Marquardt least-squares minimization) until the best fit to the observations is achieved. To exclude the locations where emission bands occur, the optimization is performed only from 1.30 -- 1.80 $\mu$m, 1.95 -- 2.50 $\mu$m, 2.91 -- 3.20 $\mu$m, 3.60 -- 4.10 $\mu$m, and 4.45 -- 4.70 $\mu$m. The particle diameters of both components in the mixture ($<D_{H_{2}O}>$ and $<D_{AC}>$) are assumed equal to 1 $\mu$m (see next Section). The modeling is applied to each pixel and the results are averaged in a 3-by-3 pixel box. Figures \ref{continuum_modeling} and  \ref{continuum_modeling1} show the comparison between the observed reflectance spectra (triangles) extracted in boxes A and B from 1.2 -- 4.8 $\mu$m and the best-fit model at same spectral resolution as the HRI-IR observations. The synthetic spectra match all the main features of the observed spectra from 1.2 -- 4.8 $\mu$m. The parameters of the best-fit models shown in Figures \ref{continuum_modeling} and \ref{continuum_modeling1} are listed in Table \ref{tabB} (see Model 1, Box A and B). Specifically, the free parameters listed in Table \ref{tabB} are the 3-by-3 pixel box averages of the modeled results together with the standard error of the mean. The same approach is applied to the reduced $\chi_{red}^{2}$. The latter in case of Model 1 computed in boxes A and B is 10 and 16, respectively, for a number of degrees of freedom equal to 230. We attribute the high value of  $\chi_{red}^{2}$ to a possible underestimation of the error and to the high quality of the data with respect to the optical constants. Furthermore, the possible presence of yet unidentified subtle emission bands in wavelength regions where the modeling is computed may raise the value of $\chi_{red}^{2}$. Additionally, the combination of our simple model and the chosen set of optical constants is not expected to fit the spectra exactly, given the many complexities of cometary comae.

The coma temperatures, $T_{c}$, derived from the modeling in both water ice-rich and water ice-depleted areas are similar. This indicates that the thermal component of the spectra is not sensitive to the water ice content and that $T_{c}$, around 300 K, represents the effective dust temperature. The scattering ($f$) and emitting ($f_{e}$) filling factors differ by a
factor of 2.0 in water ice-depleted areas (Box A, Table \ref{tabB}).  This
ratio would be increased to a factor of 3.4, if we were to adopt the
comet dust phase function of \citet{schleicher11}.  However, our
$f_{e}$ is dependent on the effective temperature of the spectrum, which
may not be the true temperature of the thermally emitting grains, the
latter being the more relevant quantity.  Moreover, the thermal
emission spectrum may be the result of the superposition of many
different grains with different radiative temperatures.  Finally, the
grains that dominate the scattered light are not necessarily the same
grains that dominate the thermal emission at these wavelengths.
Altogether, we report our best-fit values, but do not interpret them
any further.


Modeling of the 3-$\mu$m region for those spectra displaying water ice absorptions requires special care. The long wavelength shoulder of the 2.7-$\mu$m water vapor emission is partially superimposed on the water ice absorption. Therefore, it is difficult to extract information about the width of the 3-$\mu$m water ice absorption. A correct estimate of the water ice temperature and grain size relies on the knowledge of the width of the 3-$\mu$m water ice absorption as discussed in Sections \ref{Particle Grain Size} and \ref{Water Ice Temperature}. As a means of evaluating our model, we compare the width of the water gas emission band in the emissivity spectra extracted from both water ice-rich and water ice-depleted regions. For this purpose we avoid areas in the cometary coma which display the densest enrichment of water gas (\textit{e.g.}, the water gas plume near the waist of the nucleus discussed by \citet{Hearn2011}), where optical depth effects are expected and may change the shape and width of the water gas emission. Figure  \ref{continuum_modeling_radiance} , panels (a) and (b), show the same spectra discussed above, but
converted to radiance to allow inspection of the emission bands. Panels (c) and
(d) show the resultant emissivity spectra obtained by subtracting the model
from the observed radiance spectra.
The emissivity spectra outside the emission bands are close to zero, indicating that the continuum is adequately  modeled as well as the width and the depth of the 1.5- and 2.0-$\mu$m water ice absorptions. The widths of the water gas emission extracted in both depleted
and rich water ice regions are comparable extending from 2.55 to 2.90 $\mu$m, indicating that the 3.0-$\mu$m water ice absorption band has
been modeled correctly, thus validating our approach.

\subsection{Water Ice and Refractory
Grains: Particle Size Effects }\label{Particle Grain Size}
Figure \ref{GrainSizesrefr} shows the comparison between the water ice-depleted spectrum observed 7 min post-CA and the synthetic spectrum obtained from an areal mixture of water ice ($<D_{H_{2}O}>$ =  1 $\mu$m) and amorphous carbon, with $<D_{AC}>$ equal to 1 $\mu$m, 2 $\mu$m, and 5 $\mu$m. The modeling results are listed in Table \ref{tabB} (see Box A, Model 1, 2, and 3, respectively). The contribution of water ice is always less than 0.1\%, as expected since the observed spectrum is featureless. The modeling analysis of the water ice-depleted spectra is therefore only sensitive to the characteristics of the refractories, allowing us to investigate the particle size effects of the refractory grains. The observed spectral slope is compatible with amorphous carbon on the order of 1 $\mu$m. However, because amorphous carbon is strongly absorbing, its calculated reflectance is not strongly dependent on grain size, as indicated by each model's $\chi_{red}^{2}$ (Table \ref{tabB}). The results from this analysis support holding the refractory grain size equal to 1 $\mu$m in the next paragraph.

Figure \ref{GrainSizes} shows the comparison between the water ice-rich reflectance spectrum of box B in Figure \ref{errorRadiance} and the synthetic spectrum obtained from an areal mixture of amorphous carbon ($<D_{AC}>$ =
1 $\mu$m) and water ice, with $<D_{H_{2}O}>$ as a free parameter. Results are given in Table \ref{tabB} (Box B, Model 2). The best value for $<D_{H_{2}O}>$ (red line) is on the order of 1 $\mu$m (0.82 $\mu$m) and the goodness of fit is the same as in the case of Model 1 obtained with a fixed value of 1 $\mu$m for $<D_{H_{2}O}>$. Water ice particles on the order of 2 $\mu$m in size provide a similarly good fit to the data (see blue line, Table \ref{tabB}, Box B, Model 3). The purple, green, and brown lines in Figure \ref{GrainSizes} represent the synthetic spectra for $<D_{H_{2}O}>$ equal to 5, 10, and 100 $\mu$m, respectively.  The fits in these cases are considerably worse, as also indicated by their $\chi_{red}^{2}$. This is because the width and relative strength of the water ice absorption bands at 1.5, 2.0, and 3.0 $\mu$m are strongly dependent on the water ice grain size, as shown in the inserted panel of Figure \ref{GrainSizes}. We therefore conclude that the three absorption features observed at 1.5, 2.0, and 3.0 $\mu$m are consistent in bandwidth and strength with the presence of water ice grains of size less than 5 $\mu$m in the coma. We acknowledge that even if $<D_{H_{2}O}>$ is generally smaller than $D_{H_{2}O}$, which is what appears in the $X=\pi D/\lambda >> 1$ relation, we are not necessarily comfortably within the \citet{Hapke1993} geometric optics regime.

\subsection{Phase and Temperature of the Water Ice}\label{Water Ice Temperature}
Laboratory measurements show that infrared water ice absorption bands change position and shape
as a function of phase (crystalline or amorphous) and temperature \citep{Grundy1998,Mastrapa2008}. Figure \ref{temperature_effects} illustrates the variations in the modeling of the Hartley 2 water ice-rich spectrum for different phases and temperatures of the water ice. Specifically, we compare the spectral modeling obtained using \citet{Mastrapa2009} optical constants of amorphous (Ia)  and crystalline (Ic) water ice at 120 K (purple and yellow lines in Figure \ref{temperature_effects} and Models 7 and 8 in Table  \ref{tabB}, respectively) and \citet{Warren2008} optical constants of crystalline water ice at 266 K (red line in Figure \ref{temperature_effects}, Model 1 Box B in Table \ref{tabB}).
The inserted panel in Figure \ref{temperature_effects} shows the comparison between reflectance spectra of 100\% water ice on the order of 1 $\mu$m computed using these three sets of optical constants. 

The 1.65-$\mu$m water ice absorption band is strongly dependent on phase and temperature (see Figure \ref{temperature_effects}, inserted panel). This band is evident in reflectance spectra of cold crystalline ice \citep{Grundy1998}, but less prominent or absent in amorphous ice. Crystalline ice indicates formation temperatures in excess of 130 K, the critical temperature for transformation from amorphous to crystalline ice \citep{Grundy1998,Jewitt2004}. However, the 1.65-$\mu$m band reduces in strength as temperature increases, and
it almost disappears in crystalline ice T$\geq$ 230K \citep{Grundy1998,Mastrapa2008}. We do not detect the 1.65-$\mu$m absorption band in Hartley 2 spectra, so we can not distinguish between amorphous water ice and ``warm" crystalline water ice.  

Another band strongly dependent on the water ice deposition temperature is the 3-$\mu$m band. As observed by \citet{Mastrapa2009}, the band at 3 $\mu$m is stronger and shifted to longer wavelengths in crystalline water ice compared to that of amorphous water ice. The differences between the \citet{Mastrapa2009} and \citet{Warren2008} optical constants around 3 $\mu$m, discussed by \citet{Mastrapa2009}, are possibly due to differences in laboratory methods (see Figure \ref{temperature_effects}, inserted panel). Qualitatively, the \citet{Warren2008} data set provides a better fit to the data around 3.5 $\mu$m. However, comparing the modeling results reported in Table \ref{tabB}, we can not determine the phase (amorphous versus crystalline) and temperature of the water ice. A limiting factor is the spectral differences seen in laboratory measurements between amorphous and crystalline ice around 3 $\mu$m fall in regions overlapping with gas emissions, which are not used to optimize our modeling.
\subsection{Well Separated or Intimate Mixing}\label{Areal vs Intimate}
Using a linear mixing model, the Hartley 2 data are consistent with 1-$\mu$m diameter water ice particles. The resulting model, shown in Figure  \ref{GrainSizes}, matches the ice absorptions
present in the data at 1.5, 2.0 and 3.0 $\mu$m (Section \ref{Particle Grain Size}). Figure \ref{areal_intimate} shows that no particle size of
water ice is consistent with the observations when intimately mixed with refractory materials. The results displayed in
Figure \ref{areal_intimate} are from models of water ice and amorphous carbon ($<D_{AC}>$ =
1 $\mu$m) using
optical constants from \citet{Warren2008} and \citet{Edoh1983}, respectively.  The free parameters in our model are the water ice-to-dust ratio ($F_{H_{2}O}$), the coma temperature $T_{c}$, and the scattering ($f$) and emitting ($f_{e}$) filling factors. Large water ice particle sizes (\textit{e.g.}, 10 $\mu$m) match the ice absorption
near 2 $\mu$m, but the  3-$\mu$m region is modeled poorly. When using smaller particle sizes, the models have weak or absent 2-$\mu$m
bands with respect to the observations and 3-$\mu$m absorptions that are narrower than the data. We can therefore conclude that intimate mixing can be excluded.  The fact that the observed spectra, particularly the strengths and shapes of the three absorption bands of water ice,  are well reproduced by an areal mixture of water ice and refractories implies that water ice
in the coma of Hartley 2 is relatively pure.

\section{Spatial distribution of the water ice and refractories}
The modeling analysis described in Sections \ref{Modeling} and \ref{Results} (areal mixture of water ice and amorphous carbon) is applied to each pixel of the  5006000 and 5007002 scans. Over the entire scans, no variation in particle size was detected. Thus, for each scan we compute the water ice-to-dust ratio map ($F_{H_{2}O}$) and the scattering filling factor map ($f$), assuming a grain size of 1 $\mu$m for water ice and dust. We compute the water ice and dust column density as
\begin{equation}\label{eq_column_density}
N_{j} \leq \frac{F_{j}f}{\pi (<D_{j}>/2)^2}
\end{equation}
where $j$ refers to water ice or dust (represented by amorphous carbon). In the case of dust, we have $F_{j} = 1-F_{H_{2}O}$. The column density of the water ice and dust, expressed in number of particles per square meter, is shown in Figure \ref{column density}. These maps have been smoothed applying a resistant mean (2.5 $\sigma$ threshold) in a 3-by-3 pixel box. The column density, $N_{j}$, computed via Equation \ref{eq_column_density}, is an upper limit, given that $<D_{j}>$ represents the diameter of the particles within a grain (see Section \ref{Modeling}), not the total grain aggregate diameter. 

Inspection of panel (B) of Figure \ref{column density} reveals that water ice is not uniformly distributed in the innermost coma of Hartley 2. It is possible to observe the presence of water ice-enriched ($\geq$6$\times$10$^{6}$ particles/m$^{2}$, up to values of $\sim 2.5\times10^{9}$ particles/m$^{2}$ near the limb of the small lobe of the nucleus facing the Sun) and water ice-depleted ($\leq$ 10$^{6}$ particles/m$^{2}$) regions. Water ice is found in 5 out of the 6 strongest jets (J$_{1}$ through J$_{5}$, as labeled in panel A). The only jet not enriched in water ice is the one off the side of the large lobe facing the Sun and labeled in panel (A) as J$_{6}$. The comparison between the HRI-VIS context image acquired 7 min post-CA (panel A) and the corresponding water ice column density map (panel B) suggests a strong correlation between the brightness distribution in the coma of Hartley 2 and the water ice abundance, as seen by \citet{Hearn2011}. Unlike water ice, dust is everywhere in the innermost coma of Hartley 2, with an enhancement in the jets (Figure \ref{column density}, panel C). Because the water ice and dust column densities differ by a factor of 10, we can conclude that the innermost coma of Hartley 2 is dominated by dust. Similar conclusions are obtained for the water ice and dust column density maps derived from the 5007002 scan acquired 23 min post-CA, although the spatial resolution is lower than scan 5006000 (Figure \ref{column density}, right column). As observed in Section \ref{Results}, amorphous carbon is one of the lowest albedo reddening agents, therefore the dust column density provided here represents an upper limit. However, even if astronomical silicates are used as a refractory component instead of amorphous carbon (not shown), the innermost coma of Hartley 2 is still dominated by dust. 
\subsection{Driver of activity}
Figure \ref{activity} shows the spatial distribution of the water ice and dust in the innermost coma of Hartley 2 compared with that of gaseous water and CO$_{2}$, as observed 7 min (ID scan 5006000, left column) and 23 min (ID scan 5007002, right column) post-CA. The gaseous-water and CO$_{2}$ maps have been obtained computing the total of the emissivity spectral cubes (see Section \ref{Results}, Figure \ref{continuum_modeling_radiance}) in the 2.55 -- 2.90 $\mu$m and 4.15 -- 4.45 $\mu$m wavelength ranges, respectively. The maps have been smoothed by averaging in a running 3$\times$3 pixel box. The water vapor and CO$_{2}$ spatial distributions are discussed here only qualitatively, since a quantitative analysis is presented by \citet{Hearn2011} and \citet{Feaga2012}. The maps in Figure \ref{activity} show a water vapor-rich region extending
roughly perpendicular to the waist of the nucleus (panels A and E). This region has relatively little CO$_{2}$ (panels B and F) and no water ice. The CO$_{2}$ is concentrated in jets, with a particular enhancement in the region off the end of the smaller lobe of the nucleus facing the Sun (jet J$_{1}$ in panel A of Figure \ref{column density}). The
spatial distribution of the water ice grains (panels D and H in Figure \ref{activity}) and dust (panels C and G in Figure \ref{activity}), very different than the water vapor distribution, is
strongly correlated with the CO$_{2}$-rich jets (most strongly in the jet J$_{1}$). This correlation validates the suggestion put forth by \citet{Hearn2011} that
CO$_{2}$, rather than water gas, drags water ice grains into the coma as it leaves the nucleus. Our improved spatial distribution maps of water ice and CO$_{2}$, including additional temporal coverage and maps of the dust distribution, reinforce the importance of CO$_{2}$ in Hartley 2's activity not only for the water ice but also for the dust. 

\subsection{Radial Distributions of Water Ice and Dust}\label{Radial Distributions of Water Ice and Dust}
We focus on the variations of the water ice and dust column densities with nucleocenteric distance $\rho$ along the jets J$_{1}$ and J$_{4}$ observed at CA+7 min (see panel A of Figure \ref{column density}), with the goal of shedding light on the mechanisms that govern the inner coma of Hartley 2. Our analysis is conducted in the J$_{1}$ and J$_{4}$ jets since they are the brightest. We compute the average column densities in annular sectors with radii $\rho$ and $\rho$+d$\rho$, with d$\rho$ = 0.1 km, and azimuthal angles defined by the dash-dot green lines in Figure \ref{column density}, panel (A), which enclose the area of the jets J$_{1}$ and J$_{4}$. For the purposes of calculation, we assume the source of the jets is the center of the surface of curvature at which the material leaves the nucleus. Therefore, $\rho$ specifically represents the distance from the center of curvature. The variation of column density with $\rho$ can suggest whether a particular component sublimes, fragments, is accelerated, or is in a constant outflow. For a ÒnormalÓ cometary coma described
by a simple outflow, the column density would follow a 1/$\rho$ radial profile. Jets and fans do not change this trend. Figure \ref{profiles} shows the column density of dust and water ice as a function of $\rho$ in jets J$_{1}$ off the small lobe facing the Sun and J$_{4}$ off the lower edge of the larger lobe in the anti-sunward direction. Both dust profiles can be well fitted by the function
\begin{equation}
f(\rho)=\alpha\rho^{k}.
\end{equation}
The modeling of the dust profiles is presented in Figure \ref{profiles} and the values of $\alpha$ and $k$ are given in Table \ref{tab_profiles}. Since $k$ is close to -1, we conclude that the dust presents a constant outflow profile. Water ice presents a steeper profile than $1/\rho$, which suggests that water ice behaves differently than the dust and possibly that it sublimes very close to the nucleus in the coma. A steeper profile than $1/\rho$ could also be explained in terms of water ice particles being accelerated. However, there is no reason for which the dust would not be accelerated as well. We therefore favor the sublimation scenario. Following the approach of \citet{Tozzi2004}, we satisfactorily fit the water ice profiles with an exponential function
\begin{equation}\label{equation_profiles}
f(\rho)=\frac{\beta}{\rho}e^{-\frac{{\rho-\rho^{\prime}}}{L}}
\end{equation}
where $\rho^{\prime}$ is the distance  from the center of curvature to the surface and $L$ is the scale-length. This relationship reproduces the observed profiles (see dashed line in Figure \ref{profiles}). The values for $\beta$, $\rho^{\prime}$, and $L$, used to generate the best fit are reported in Table \ref{tab_profiles}. Given that the jet J$_{1}$ is propagating along the sunward direction, nearly perpendicular to the line of sight, viewing geometry effects are, to a first approximation, negligible. This does not appear to be the case for the jet J$_{4}$.  The scale-length of the subliming water ice (636 m in the J$_{1}$ jet) is compatible with lifetimes of 1-$\mu$m pure water ice grains at 1 AU \citep[$\sim$1 h, corresponding to three times the $e$-folding time; ][]{Hanner1981} for velocities on the order of 0.5 m/s. 

\subsection{Water Ice Mass Production Rate}
The modeling of the water ice column density radial profile presented in the previous section can be used to estimate the water ice mass production rate. For the same arguments reported above, we focus on the jet off the small lobe of the nucleus facing the Sun (jet J$_{1}$). The production rate $Q$ of a material with a $\Gamma/\rho$ column density profile within a jet of angular size $\theta$ is given by:
\begin{equation}\label{production rate} 
Q = \frac{2\:\pi\:\Gamma\:v\:(1-cos\frac{\theta}{2})}{{\theta}},
\end{equation}
where $v$ represents the outflow speed. With Equations \ref{equation_profiles} and \ref{production rate}, we can compute the water ice production rate. Letting $v=0.5$ m/s and $\Gamma=\beta$, we obtain in the jet J$_{1}$ ($\theta = 54^\circ$) a water ice production rate $Q_{H_{2}O-ice}$ of $9\times10^{11}\pm2\times10^{10}$ particles/s, which corresponds, for 1-$\mu$m particles with density of 1 g cm$^{-3}$,  to a mass production rate of $\dot{M} = 0.00046\pm0.00001$ kg/s. The 2\% error on $Q_{H_{2}O-ice}$ and $\dot{M}$, has been computed considering the errors on $\beta$ and $L$, reported in Table \ref{tab_profiles}. However, according to our model the 1-$\mu$m grains are in
aggregates, and as such, their interiors may be shielded from our
view.  If so, and if the interiors are also comprised of water ice,
our derived mass production rate is necessarily a lower limit.

\section{Discussion}
Comet Hartley 2 is a hyperactive comet with an active fraction near 1.7 -- 2.5 \citep{Kelley2013}. The presence of an icy grain halo, suggested by \citet{Lisse2009}, to account for the high water production rate of comet Hartley 2 seems plausible, but difficult to justify quantitatively. Two relevant particle populations have been detected so far in the coma of Hartley 2: (1) 1-$\mu$m water ice grains in the inner
few kilometers of the Hartley 2 coma (this paper), and (2) distinct, isolated grains that
in visible images envelope the nucleus of Hartley 2 up to 40 km, and estimated by their brightness to be centimeter-sized (1 cm in radius)
or larger and possibly made of ice (density of 0.1 g/cm$^{3}$) \citep{Hearn2011,Kelley2013}. If the latter particles are indeed icy, both will contribute to the total water production rate \citep[\textit{e.g.}, 10$^{28}$ molec/s $\simeq$ 299 kg/s][]{Hearn2011} up to 2$\times$10$^{-4}$\% (upper limit estimated considering the 2\% error on $\dot{M}$) and 0.5\% \citep{Kelley2013}, respectively. Therefore, the 1-$\mu$m water ice particles and the cm particles observed in the visible images do not appear to be the source of the comet's enhanced water production. These observations, collectively, raise interesting questions on the relationship between the 1-$\mu$m and cm size particles and the source of the hyperactivity of comet Hartley 2.

The estimated velocity consistent with our derived scale-lengths, 0.5 m/s, is much lower than the expansion velocities expected for isolated 1-$\mu$m water ice grains, the latter being on the order of 600 m/s at 1 AU \citep{Hanner1981,Whipple2}. This incongruency suggests that the water ice is in the form of large aggregates (millimeter to centimeter size), consistent with the applied spectroscopic modeling and the low inferred velocities. Our suggestion is based on the modeling by \citet{Fougere2013}, who predict velocities up to 100 m/s and 2 m/s for grains of 1 $\mu$m and 5 mm, respectively. If we consider the case of large aggregates, \textit{e.g.}, particles on the order of 1 cm in radius, with density of 0.1 g cm$^{-3}$ and our derived velocities of 0.5 m/s, we estimate the aggregate contribution to the total water production rate to be 0.3\%, in agreement with the favorable scenario put forth by \citet{Kelley2013}. This suggests that the large chunks observed in the visible are most likely made of 1-$\mu$m water ice particles. One caveat important to mention is that the expansion velocity value of 0.5 m/s, has been measured considering the $e$-folding time of 1-$\mu$m water ice particles at 1 AU reported by \citet{Hanner1981}, which refers to single particles and not aggregates. Further studies are needed to improve our understanding of the lifetimes of water ice aggregates. 

The conclusion that the 1-cm large chunks detected in the visible data are aggregates of 1-$\mu$m water ice particles is supported by our estimate of the total water ice cross section, given by $f\:F_{H_{2}O-ice}\:\sigma$, being $\sigma$ the geometric cross section of one pixel, and equal to 0.001 km$^{2}$ in a field of view of 4.7$\times$2.7 km$^{2}$ as measured from the HRI-IR data acquired 7 min and 23 min post-CA. This value is in agreement with the total icy particle cross section estimated from the MRI observations of the large chunks being in the range of 0.0004-0.003 km$^{2}$ within 20.6 km from the nucleus \citep{Kelley2013}.

Even if, at first order approximation, we are able to reconcile our analysis with the MRI-and HRI-VIS observations of the large chunks in the coma of Hartley 2, we still find some discrepancies when comparing our results with other observations. The measured scale-length of $\sim$600 m implies that not much water ice would survive up to 40 km from the nucleus, in contradiction with the MRI-VIS observations of the large chunks up to this distance \citep{Kelley2013}. If water ice particles were entrained in the CO$_{2}$ gas flow, as suggested by the spatial correlation between the water ice and CO$_{2}$ and assumed to be three orders of magnitude faster, then ice particles could be observed at larger distances from the nucleus. However, not only ice but also dust would be carried by the CO$_{2}$ and no evidence of acceleration is seen in the dust profiles. Our observations are still unable to solve the mystery about the main source of water in Hartley 2's  coma. Indeed, while we can rule out the case of isolated 1-$\mu$m water ice grains as the main contributor to the total water production rate, the case of aggregates of 1-$\mu$m remains open. To confirm or reject the prediction by \citet{Fougere2013} that the sublimating icy grains in the Hartley 2's coma release $\sim$77\% of the total water particles, advanced studies are needed to further understand how water ice aggregates sublime and constrain the unknown size and density of the aggregates. We have so far considered as size and density of these aggregates the values suggested by \citet{Kelley2013} to describe the large chunks observed in the MRI- and HRI-VIS data. However, a final resolution on the relationship between 1-$\mu$m water ice grains detected in near-IR scans and the large chunks described by \citet{Kelley2013} is still open. We have recently identified in near-IR scans point sources corresponding to the large chunks observed in HRI- and MRI-VIS data. The analysis of the spectroscopic properties of these point sources, which will be subject of a future work, will shed light on the composition of the large chunks observed in the visible data together with their relationship with the 1-$\mu$m water ice grains detected in the near-IR scans, and maybe help us solving the mystery about the main source of water in Hartley 2's  coma.

This paper focused on producing the best dataset yet available for studying
a comet's innermost coma.  We also included some very basic modeling, with
several of simplifying assumptions, to get some preliminary results regarding
the comet's environment.  The discrepancies that we find between these
results and other observations show that these simple models are clearly
insufficient for explaining the data, and detailed models must be
developed to produce a more comprehensive explanation of the observations.

\section{Conclusions}
We have analyzed DI HRI-IR spatially resolved scans of comet Hartley 2, collected 7 min (ID 5006000) and 23 min (ID 5007002) after closest approach, when the comet was at a heliocentric distance of 1.06 AU and with spatial scales at mid-scan of 55 m/pix and 173 m/pix, respectively.  The reflectance spectra show clear absorption bands near 1.5, 2.0, and 3.0 $\mu$m, consistent in central wavelength, band width, and band shape with the presence of water ice grains in the coma.  Using Hapke's radiative transfer model we find that the spectra are best fit by an areal mixture of a featureless, highly absorbing, refractory component and water ice grains with size on the order of 1 $\mu$m.  Intimate mixtures of the same materials cannot reproduce the spectra; a strong indication that the grains of water ice are relatively pure, \textit{i.e.}, relatively free of refractory material. In addition, the measured temperature of the grains is on the order of 300 K, which is inconsistent with water ice persisting long enough to be observed, implying that the thermal emission is purely from the refractory component and that the refractory grains are not in thermal contact with the grains of water ice. 

In the innermost coma, within a nuclear radius of the surface, dust dominates over water ice by a factor 10, with column densities of 10$^{9}$ and 10$^{10}$ particles/m$^{2}$ of water ice and dust, respectively.  The water ice and the dust are not uniformly distributed -- their spatial distribution is strongly correlated with the presence of CO$_{2}$ emission but very different from the distribution of water vapor emission, indicating that the CO$_{2}$ plays a key role in Hartley 2's activity for both water ice and dust and that the ``waist jet'' of gaseous H$_{2}$O arises from a different process than most of the outgassing.  The radial profiles of the refractory dust follow a 1/$\rho$ profile, consistent with uniform outflow, while the radial profiles of water ice are much steeper.  The differences between the profiles could be consistent with acceleration of the water ice but not the dust. However, it is more likely that the steeper profiles for ice are due to water ice sublimation.  The lifetime of water ice grains at 1 AU \citep{Hanner1981} suggests that the expansion velocity of the water ice grains is of order 0.5 m/s.  This in turn suggests that the 1-$\mu$m grains seen in the spectra are the components of more massive aggregates, which could imply shielding of some grains even in the optically thin coma and that the column densities of water ice are higher than deduced from our modeling.

One can compare the icy grains in comet Hartley 2, a hyperactive comet from which icy grains are apparently dragged out continuously by gaseous CO$_{2}$, with the icy grains mechanically excavated from 10-20-m depths in the low-activity comet Tempel 1 \citep{Sunshine2007}, and with the icy grains excavated by the infrequent but large natural outbursts of comet Holmes \citep{Yang2009}. In all three cases, the icy grains are relatively pure and dominated by particles of order 1 $\mu$m in size.  Taken together this suggests that in most comets the ice in the interior of the nuclei is in the form of aggregates of relatively pure ice and that intimate mixtures of ice and refractories are rare.  Furthermore, this implies that nuclei are commonly very porous, as suggested by the few bulk densities that have been determined and that aggregation models \citep[\textit{e.g.}, ][]{Greenberg1999} that call for grains with refractory cores, organic mantles, and icy crusts are generally inappropriate for comets.

\section*{Acknowledgments}
The authors thank Dennis Bodewits and Dennis Wellnitz for useful discussions and the anonymous referees for comments that helped improving the quality of this paper. 

This work was funded by NASA, through the Discovery Program, via contract NNM07AA99C to the University of
Maryland and task order NMO711002 to the Jet Propulsion
Laboratory.
\bibliographystyle{icarus}
\bibliography{silvia}
\newpage
\appendix\section{Scattered Flux}\label{scattered flux}

The cometary flux in the wavelength range covered by the DI HRI-IR spectrometer is the sum of the scattered ($F_{scat}$) and thermal ($F_{thermal}$) components. We describe below how to estimate $F_{scat}$ and its contribution to the reflectance.

Consider a body at a distance $r$ from the Sun, of radius $R$. The total energy incident on the surface is
\begin{equation}
L_{in}=\pi R^{2}\frac{L_{\odot}}{4 \pi r^{2}}=\frac{L_{\odot} R^{2}}{4 r^2}
\end{equation}
where $L_{\odot}$ is the luminosity of the Sun. Only part of the incident flux is scattered back. The Bond Albedo A (or spherical albedo) is defined as the ratio of the emergent luminosity to the integrated incident flux ($0\leq A \leq1$). Therefore, the luminosity is
\begin{equation}\label{*}
L_{out}=A L_{in}=\frac{A L_{\odot} R^{2}}{4 r^2}\:.
\end{equation}
The target-observer distance is $\Delta$. If radiation is scattered isotropically, the observed flux is
\begin{equation}
F_{scat}=\frac{L_{out}}{4 \pi \Delta^{2}}\:.
\end{equation}
If we assume that the reflecting object is a homogeneous sphere, the distribution of the reflected radiation only depends on the phase angle $\alpha$. Thus we can express the flux density observed at a distance $\Delta$ as
\begin{equation}\label{**}
F_{scat}=K \Phi(\alpha) \frac{L_{out}}{4 \pi \Delta^{2}}\: .
\end{equation}
The function $\Phi$ is called phase function. It is normalized so that $\Phi(\alpha=0)=1$. $K$ is a normalization constant defined such that
\begin{equation}
\int_{S}K \Phi(\alpha) \frac{L_{out}}{4 \pi \Delta^{2}} dS = L_{out}
\end{equation}
where the integration is extended over the surface of radius $\Delta$.
Substituting Eq. \ref{*} in Eq \ref{**} we have
\begin{equation}
F_{scat}=K \Phi(\alpha) \frac{A L_{\odot} R^{2}}{16 \pi \Delta^{2} r^{2}}
\end{equation}
The geometric albedo, $p$, is the ratio of the flux density of a body at phase angle $\alpha=0$ to the flux density of a perfect Lambert disk of the same radius and same distance as the body, but illuminated and observed perpendicularly. We have
\begin{equation}
p = \frac{KA}{4}\:
\end{equation}
If we express $F_{scat}$ in terms of $p$, we obtain
\begin{equation}
F_{scat}=\frac{p}{\pi} \Phi(\alpha) \frac{L_{\odot} R^{2}}{4 r^2 \Delta^{2}}\;.
\end{equation}
The solar flux at a heliocentric distance $r_{1} = 1$ AU is
\begin{equation}
F_{\odot} = \frac{L_{\odot}}{4 \pi}\;.
\end{equation}
We have then
\begin{equation}
\frac{F_{scat}}{F_{\odot}}=\frac{\Phi(\alpha) p R^{2}}{r^2 \Delta^{2}}\;.
\end{equation}
If we now consider the case of grains in a cometary coma, and we indicate with $C$ the total cross section of the cometary grains within the slit, the previous formula can be written as:
\begin{equation}\label{***}
\frac{F_{scat}}{F_{\odot}}=\frac{\Phi(\alpha)C p}{\pi r^2 \Delta^{2}}
\end{equation}
The reflectance is defined as
\begin{equation}
R = \frac{\pi I r^{2}}{F_{\odot}}
\end{equation}
where $I$ is the radiance (flux per unit solid angle).
The contribution of the scattered flux to the reflectance, indicated as $R_{scat}$, is given by
\begin{equation}
R_{scat} = \frac{\pi I_{scat} r^{2}}{F_{\odot}}= \frac{\pi F_{scat} r^{2}}{\Omega F_{\odot}}
\end{equation}
The solid angle of the coma portion we are observing is $\sigma/\Delta^{2}$, $\sigma$ being the geometric cross section. We have then
\begin{equation}
R_{scat} = \frac{\pi F_{scat} \Delta^{2}r^{2}}{\sigma F_{\odot}}
\end{equation}
Considering \ref{***}, and taking into account that the ratio between the scattering cross section and the geometric cross section is the filling factor $f$, we have
\begin{equation}
R_{scat}=f\Phi(\alpha)p
\end{equation}
\setcounter{figure}{0} \renewcommand{\thefigure}{\arabic{figure}} 
\setcounter{table}{0} \renewcommand{\thetable}{\arabic{table}} 
		\newpage		
		\begin{table}
		\caption{Characteristics of the HRI-IR scans analyzed in this paper.}
		\label{tabA}
		\centering
		\resizebox{\textwidth}{!}{%
 		\begin{tabular}{c c c}
		\hline
 		Exposure ID  & 5006000& 5007002 \\ 
 		\hline
 	Time at mid-scan at spacecraft & 2010-11-04 14:07:08 & 2010-11-04 14:23:10\\  
 	Time after CA [min] & 7 &23\\ 
 	Number of Commanded Frames & 56 & 30\\
 	Mode name & ALTFF\tmark\ & ALTFF \\   
 	Total integration time per frame [msec] & 1441 &1441\\  
 	Spatial resolution at mid-scan [m/pixel] & 55 &173\\
	Phase angle [deg]& 92 & 93 \\
	Spacecraft-comet distance [km] &5478 & 17295\\
 		\hline
		\multicolumn{2}{l}{$^{a}$The alternating binned full frame (ALTFF) stores the image in 512 $\times$ 256 pixels (spectral $\times$ spatial).}
		\end{tabular}}     
		\end{table}
		
		\begin{table}
		\caption{Characteristics of the MRI-VIS and HRI-VIS data used in this paper as context images for the IR data.}
		\label{tab_context_images}   
		\centering
		\resizebox{\textwidth}{!}{%
		\begin{tabular}{c c c c c}
		\hline  
	Exposure ID   & 5006046  & 5007065 & 5006036 & 5007064\\
		\hline
	Instrument & MRI-VIS & MRI-VIS & HRI-VIS & HRI-VIS\\
	Time (2010-11-04) & 14:06:31 & 14:25:11 & 14:06:57 & 14:24:58\\  
	Time after CA [min] & 7 &25 & 7 & 25\\ 
	Total integration time per frame [msec] & 500 & 600 & 150 & 175\\  
	Spatial resolution [m/pixel] & 50 &188 & 11 & 37\\
	Filter & CLEAR1 & CLEAR1 & CLEAR1 & CLEAR1\\
		\hline
		\end{tabular}}
		\end{table}
		
		\begin{table}
		\caption{Modeling Solutions}
		\label{tabB}   
		\centering
		\resizebox{\textwidth}{!}{%
		\begin{tabular}{c c c c c c c c c c c c}
	\hline
         	Model & $P_{H_{2}O}$$^{b}$ & $T_{H_{2}O}$ & ${H_{2}O}$-ice                & $<D_{H_{2}O}>$   & $<D_{AC}>$ & $F_{H_{2}O}$ & $T_{c}$ &$f$&$f_{e}$& $\chi^{2}_{red}$ \\
	             &                                       &  (K)                    & Optical Constants           & ($\mu$m)                 & ($\mu$m)      & (\%)                   & (K)          &     &              &                               \\
	             &                                       &                           &                                             &                                   &                        &                            &                &     &              &                               \\
	             
	\hline 
	BOX A Model1 & I$_{c}$ & 266 & \citet{Warren2008} &  1\tmark\ & 1\tmark\ & $<0.1$ & 315$\pm$3 & 0.0063$\pm$0.0001 & 0.0032$\pm$0.0006 &10.4$\pm$0.4\\

	BOX A Model2 & I$_{c}$ & 266 & \citet{Warren2008} &  1\tmark\ & 2\tmark\ & $<0.1$ & 318$\pm$3 & 0.0077$\pm$0.0002 & 0.0029$\pm$0.0006 &11.4$\pm$0.3\\
	BOX A Model3 & I$_{c}$ & 266 & \citet{Warren2008} &  1\tmark\ & 5\tmark\ & $<0.1$ & 322$\pm$3 & 0.0082$\pm$0.0002 & 0.0026$\pm$0.0005 &12.4$\pm$0.3\\

	BOX B Model1 & I$_{c}$ & 266 & \citet{Warren2008} &  1\tmark\ & 1\tmark\ & 5.1$\pm$0.2 & 306$\pm$4 &  0.0166$\pm$0.0003 & 0.0058$\pm$0.0009
& 16$\pm$1\\

	BOX B Model2 & I$_{c}$ & 266 & \citet{Warren2008} &  0.82$\pm$0.07 & 1\tmark\ & 5.3$\pm$0.2 & 307$\pm$4 &  0.0163$\pm$0.0004 & 0.0056$\pm$0.0009
& 16$\pm$1\\

BOX B Model3 & I$_{c}$ & 266 & \citet{Warren2008} &  2\tmark\  & 1\tmark\ & 5.0$\pm$0.2 & 308$\pm$4 &  0.0173$\pm$0.0004 & 0.0054$\pm$0.0009
& 17$\pm$2\\

BOX B Model4 & I$_{c}$ & 266 & \citet{Warren2008} &  5\tmark\  & 1\tmark\ & 4.9$\pm$0.2 & 311$\pm$4 &  0.0184$\pm$0.0004 & 0.0049$\pm$0.0008
& 22$\pm$2\\\

	BOX B Model5 & I$_{c}$ & 266 & \citet{Warren2008} &  10\tmark\  & 1\tmark\ & 4.7$\pm$0.1 & 313$\pm$3 &  0.0196$\pm$0.0004 & 0.0043$\pm$0.0004
& 30$\pm$3\\

	BOX B Model6 & I$_{c}$ & 266 & \citet{Warren2008} &  100\tmark\  & 1\tmark\ & 3.7$\pm$0.1 & 301$\pm$3 &  0.0255$\pm$0.0006 & 0.0063$\pm$0.0006
& 73$\pm$5\\

	BOX B Model7 & I$_{a}$ & 120 & \citet{Mastrapa2009}  &  1\tmark\  & 1\tmark\ & 5.0$\pm$0.2 & 307$\pm$4 &  0.0167$\pm$0.0003 & 0.0056$\pm$0.0009
& 16$\pm$1\\
	BOX B Model8 & I$_{c}$ & 120  & \citet{Mastrapa2009}  &  1\tmark\  & 1\tmark\ & 5.2$\pm$0.2 & 312$\pm$3 &  0.0165$\pm$0.0004 & 0.0044$\pm$0.0004
& 17$\pm$1\\
		\hline	
		\multicolumn{2}{l}{$^{a}$The particle grain sizes are fixed in the modeling}\\		
		\multicolumn{2}{l}{$^{b}$The water ice phase.}

		\end{tabular}}
		\end{table}
		
		\begin{table}	
		\caption{Values obtained from the fit of the ice and dust column density vs. $\rho$.}
		\label{tab_profiles}   
		\centering
		\resizebox{\textwidth}{!}{%
		\begin{tabular}{c c c c c c c c}
		\hline  
	Component   & function  & jet & log$_{10}$$\alpha$ & $k$ & $\rho^{\prime}$ & log$_{10}$$\beta$ &$L$\\
	  &  & & (m) &  & (m) & &(m)\\
		\hline
	Dust & $\alpha\rho^{k}$ & J$_{1}$ & 13.35$\pm$0.08 & -0.99$\pm$0.02 & &\\
	Dust & $\alpha\rho^{k}$ & J$_{4}$ & 13.07$\pm$0.04 & -1.00$\pm$0.01 & & \\
	Ice & $\frac{\beta}{\rho}e^{-\frac{{\rho-\rho^{\prime}}}{L}}$ & J$_{1}$ &  &  & 935& 12.38$\pm$0.02& 636$\pm$17\\
	Ice & $\frac{\beta}{\rho}e^{-\frac{{\rho-\rho^{\prime}}}{L}}$ & J$_{4}$ &  &  & 165& 12.12$\pm$0.05& 231$\pm$8\\
		\hline
		\end{tabular}}
		\end{table}

\newpage
\captionsetup[subfigure]{labelformat=empty}%
\begin{figure*}
\ifincludegraphics
\vspace{-30pt}
        	\centering
	\vspace{-30pt}
        	\begin{subfigure}[b]{0.4\textwidth}
        	        \caption{CA+7min}
        	        \vspace{-30pt}	
        	        \centering
        	        \includegraphics[width=\textwidth]{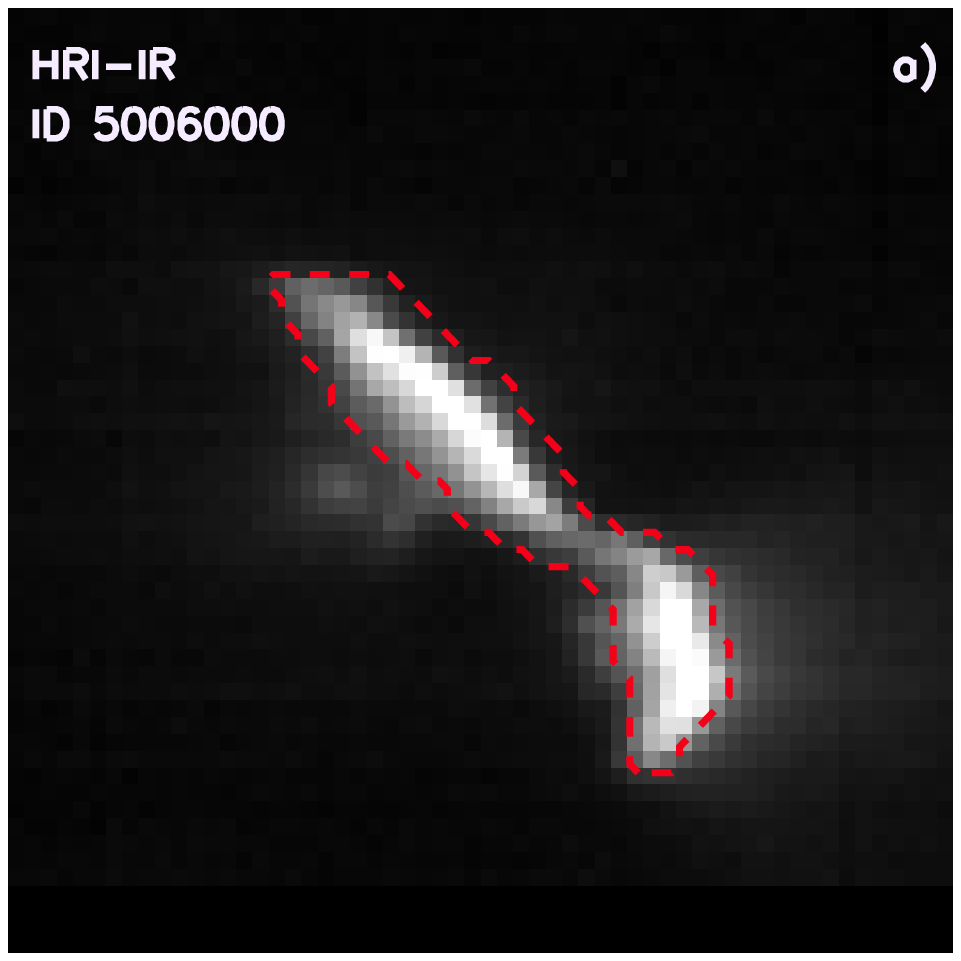}
        	\end{subfigure}
	\vspace{-30pt}
         	\begin{subfigure}[b]{0.4\textwidth}
                 \caption{CA+23min}
        	        \vspace{-30pt}	
        	        \centering
                 \includegraphics[width=\textwidth]{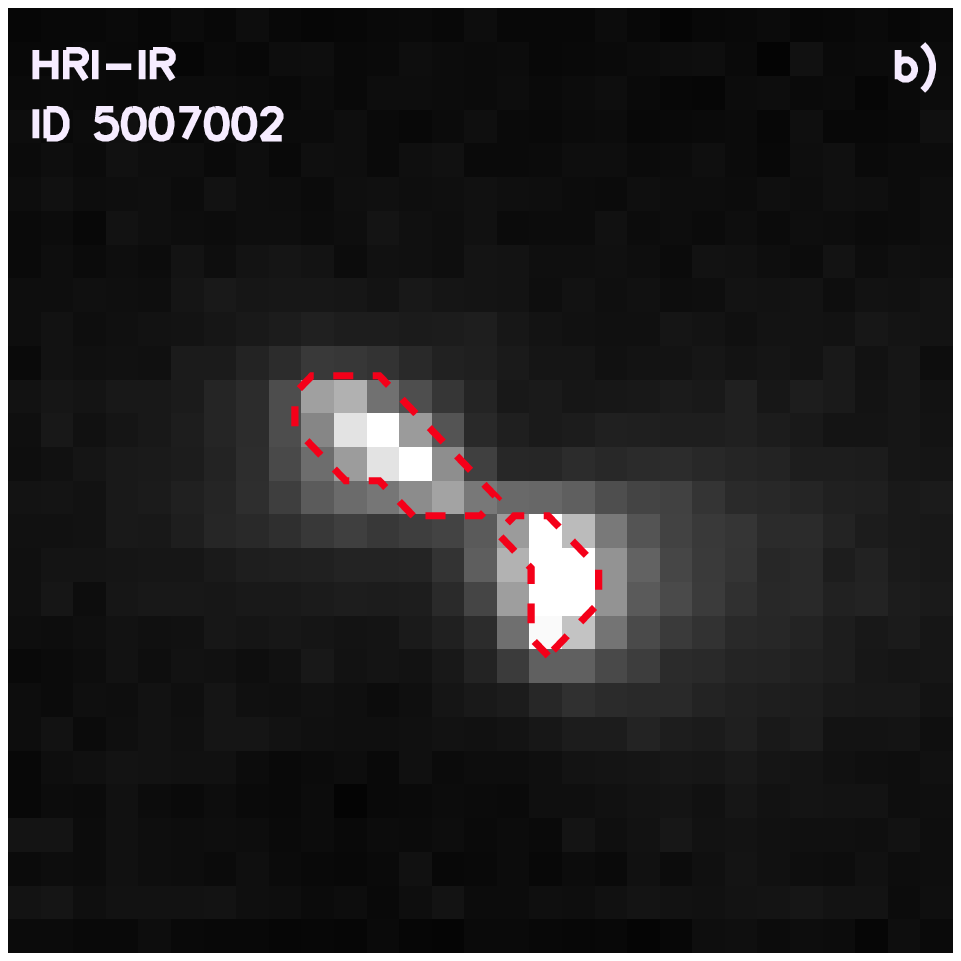}
         \end{subfigure}
         \begin{subfigure}[b]{0.4\textwidth}
        	       \centering
                \includegraphics[width=\textwidth]{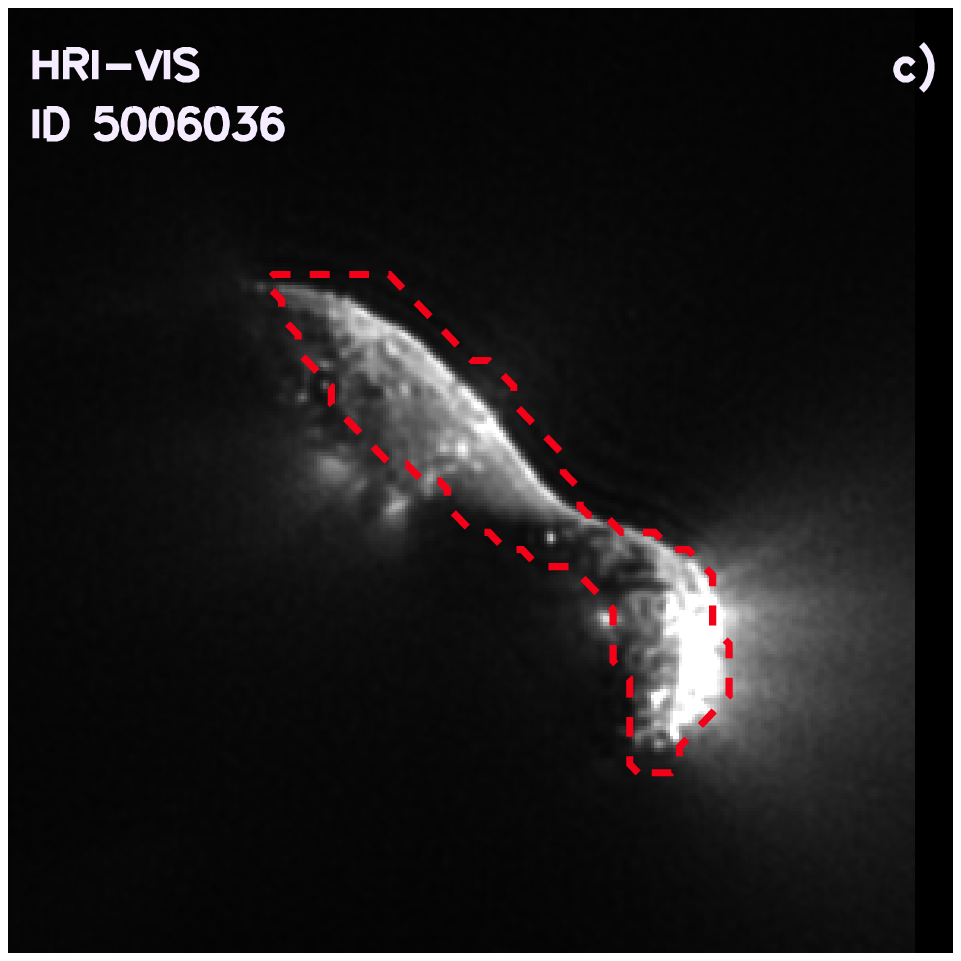}
        \end{subfigure}
        \vspace{-30pt}
        \begin{subfigure}[b]{0.4\textwidth}
        	       \centering
                \includegraphics[width=\textwidth]{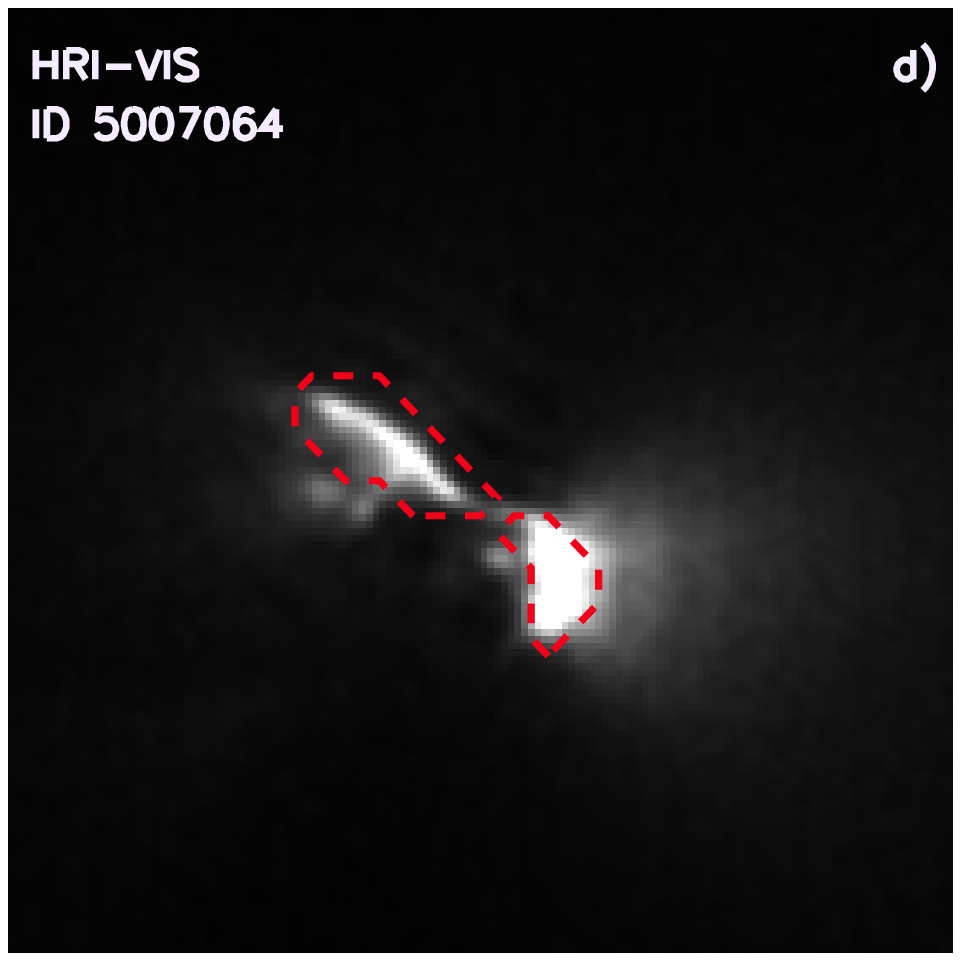}
        \end{subfigure}
        \begin{subfigure}[b]{0.4\textwidth}
        	       \centering
                \includegraphics[width=\textwidth]{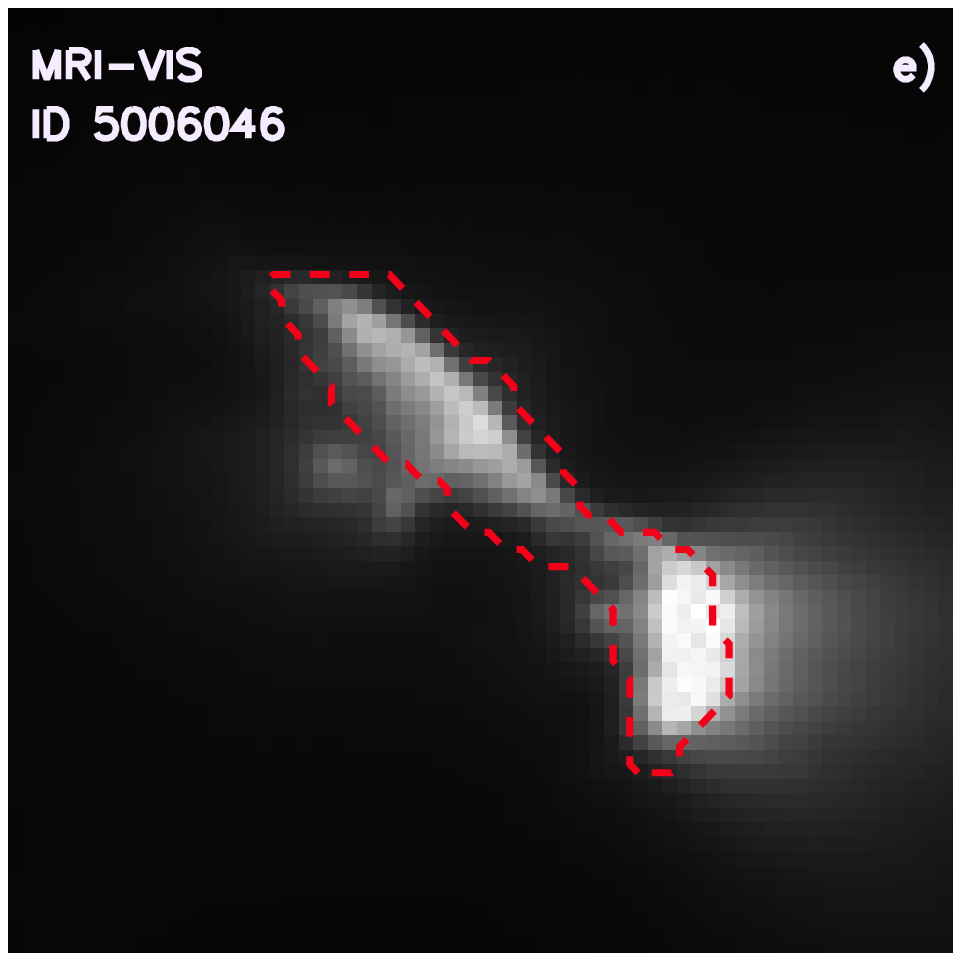}
        \end{subfigure}
        \begin{subfigure}[b]{0.4\textwidth}
        	       \centering
                \includegraphics[width=\textwidth]{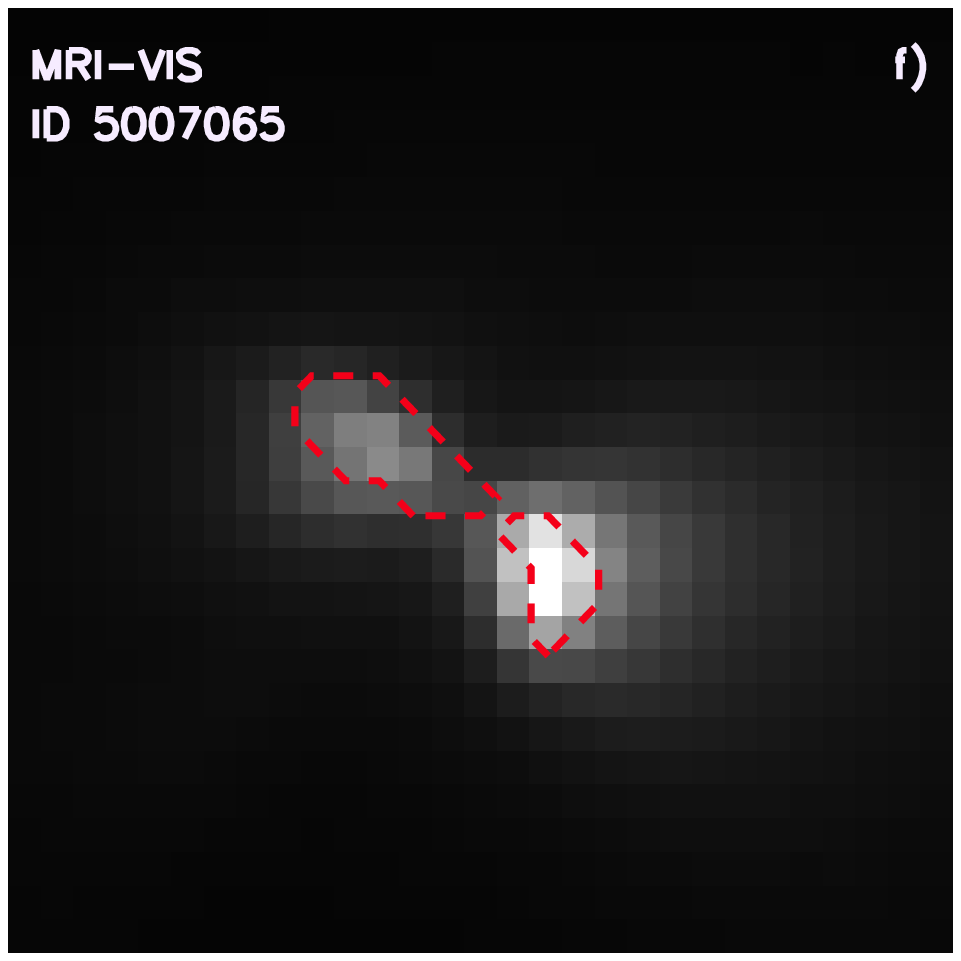}
        \end{subfigure}
        \fi
	        \caption{The HRI-IR spectral cubes acquired 7 and 23 min post-CA (see Table \ref{tabA}) at 2.5 $\mu$m are displayed in panels (a) and (b), respectively. Panels (c) and (d) show the context images of Hartley 2 taken with the high resolution camera HRI-VIS, corresponding to the ID sequences 5006036 and 5007064, respectively. The images have been deconvolved \citep{Lindler2013}. The bottom 2 panels display the MRI-VIS context images. Table \ref{tab_context_images} lists the main characteristics of the HRI-VIS and MRI-VIS context images shown in this Figure. The dashed red line contours the illuminated portion of the nucleus that is
masked for our analysis (for details see text). The Sun is to the right. \textit{See the electronic version of the Journal for a color version
of this figure.}}
	        \label{radiance-maps}
\end{figure*}

\begin{figure}
	\ifincludegraphics
	\centering	
	\includegraphics[width=12cm]{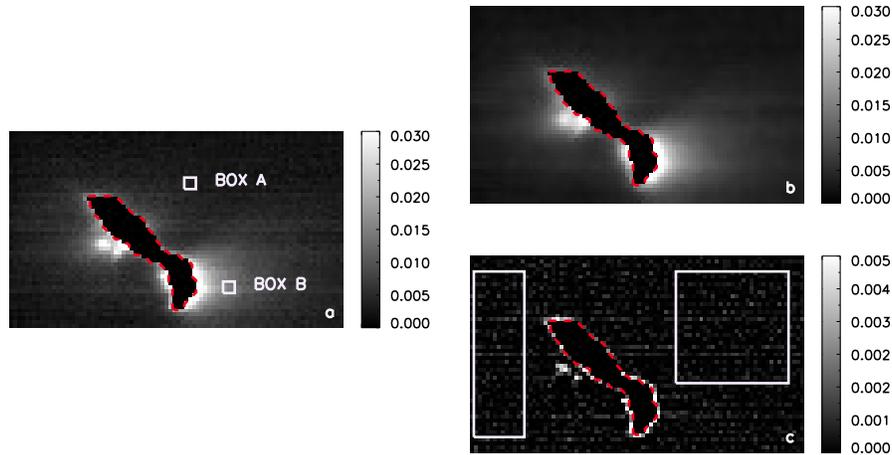}
	\fi
	\caption{Radiance error determination. (a) Radiance map acquired 7 min post-CA (ID 5006000) at 2.5 $\mu m$. The color bar indicates the line of sight radiance values (W m$^{-2}$sr$^{-1}$$\mu$m$^{-1}$). The white boxes represent the regions sampled to produce the reflectance spectra in Figure \ref{continuum_modeling}. (b) Radiance map after mean filtering with a 3$\times$3 pixel square kernel. (c) Radiance background map obtained by subtracting (b) from (a). The white boxes surround the regions used to compute the mean error (see text for details). In all maps, the nucleus has been masked and the dashed red line contours the nucleus. The Sun is to the right. \textit{See the electronic version of the Journal for a color version
of this figure.}}
        \label{errorRadiance}
\end{figure}
\begin{figure}
         \ifincludegraphics
 	\centering
	\includegraphics[width= 0.7\textwidth]{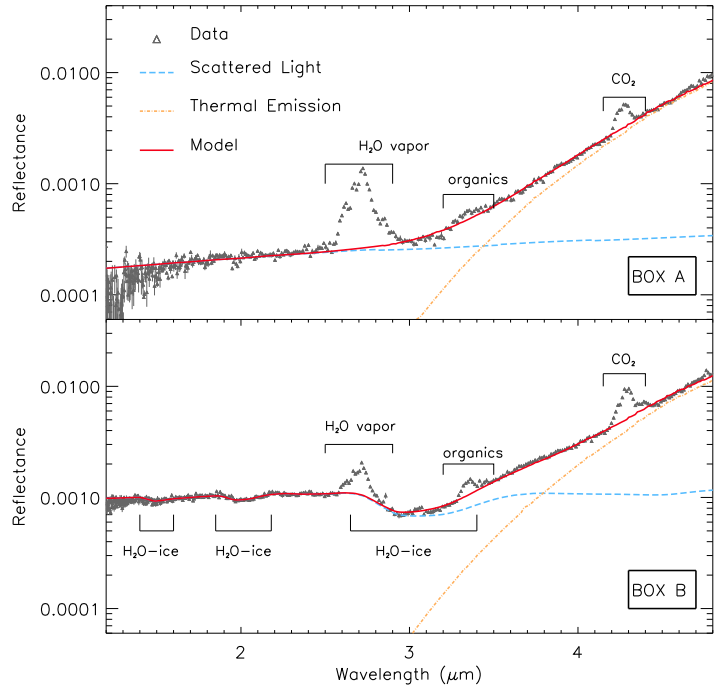}
	\fi
	\caption{Comet 103P/Hartley 2 reflectance spectra (gray triangles) over the wavelength range 1.2 -- 4.8 $\mu$m. Top and bottom panels show the spectrum (triangles) obtained performing the resistant mean (threshold 2.5$\sigma$) of the 9 reflectance spectra in box A (top panel) and B (bottom panel) of Figure \ref{errorRadiance} (panel a), respectively.  The modeling (solid red line) is the sum of the scattered solar radiation (dashed blue line) and
a blackbody function (dash-dot orange line). The spectrum shows clearly the water
vapor, organic, and carbon dioxide
emissions at 2.7 $\mu$m, 3.3 $\mu$m, and 4.3
$\mu$m, respectively, which are not fit by
the model as expected. The species responsible for the emission and absorption features are marked in the plots. \textit{See the electronic version of the Journal for a color version
of this figure.}}
	\label{continuum_modeling}%
\end{figure}
\begin{figure}
	\ifincludegraphics
 	\centering
	\includegraphics[width= \textwidth]{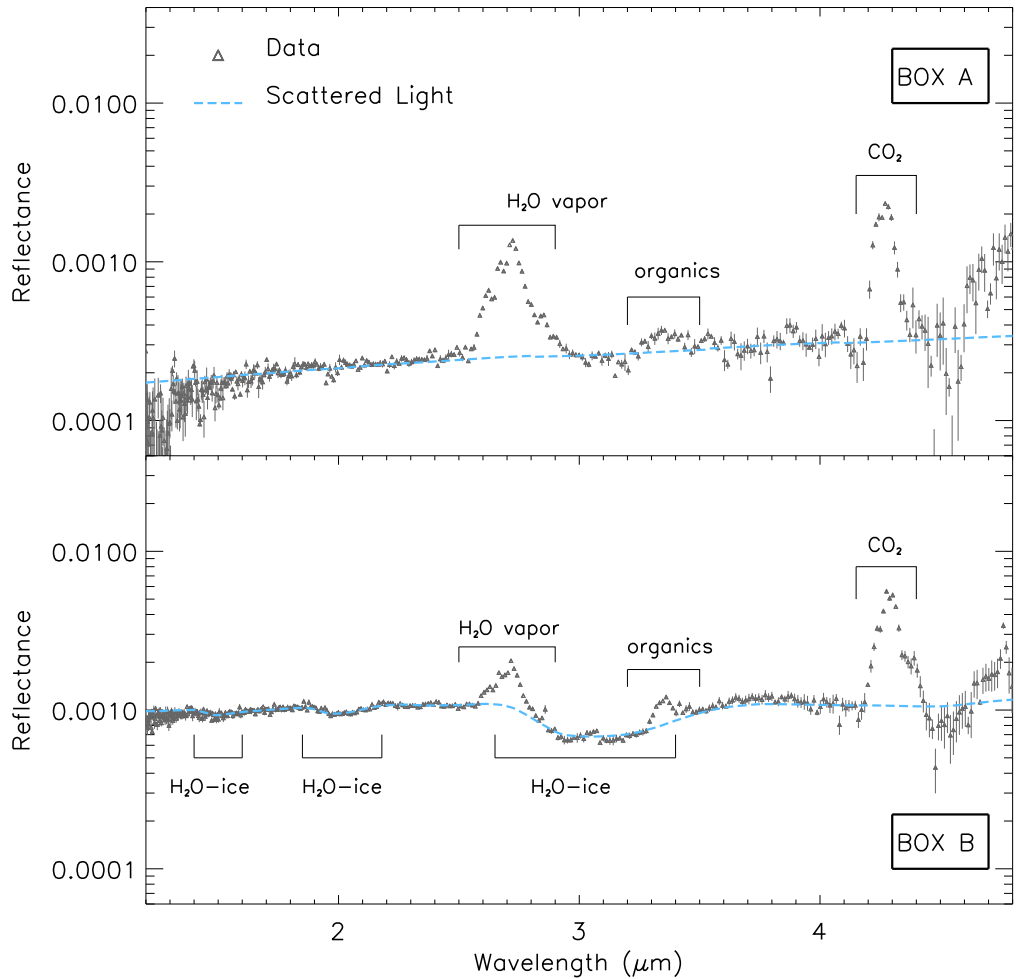}
	\fi
	\caption{Comet 103P/Hartley 2 reflectance spectra shown in Figure \ref{continuum_modeling} after the removal of the thermal models (gray triangles). Overplotted is the scattered model (dashed blue line). \textit{See the electronic version of the Journal for a color version
of this figure.}}
	\label{continuum_modeling1}%
\end{figure}
\begin{figure}
	\ifincludegraphics
 	\centering
	\includegraphics[width=\textwidth]{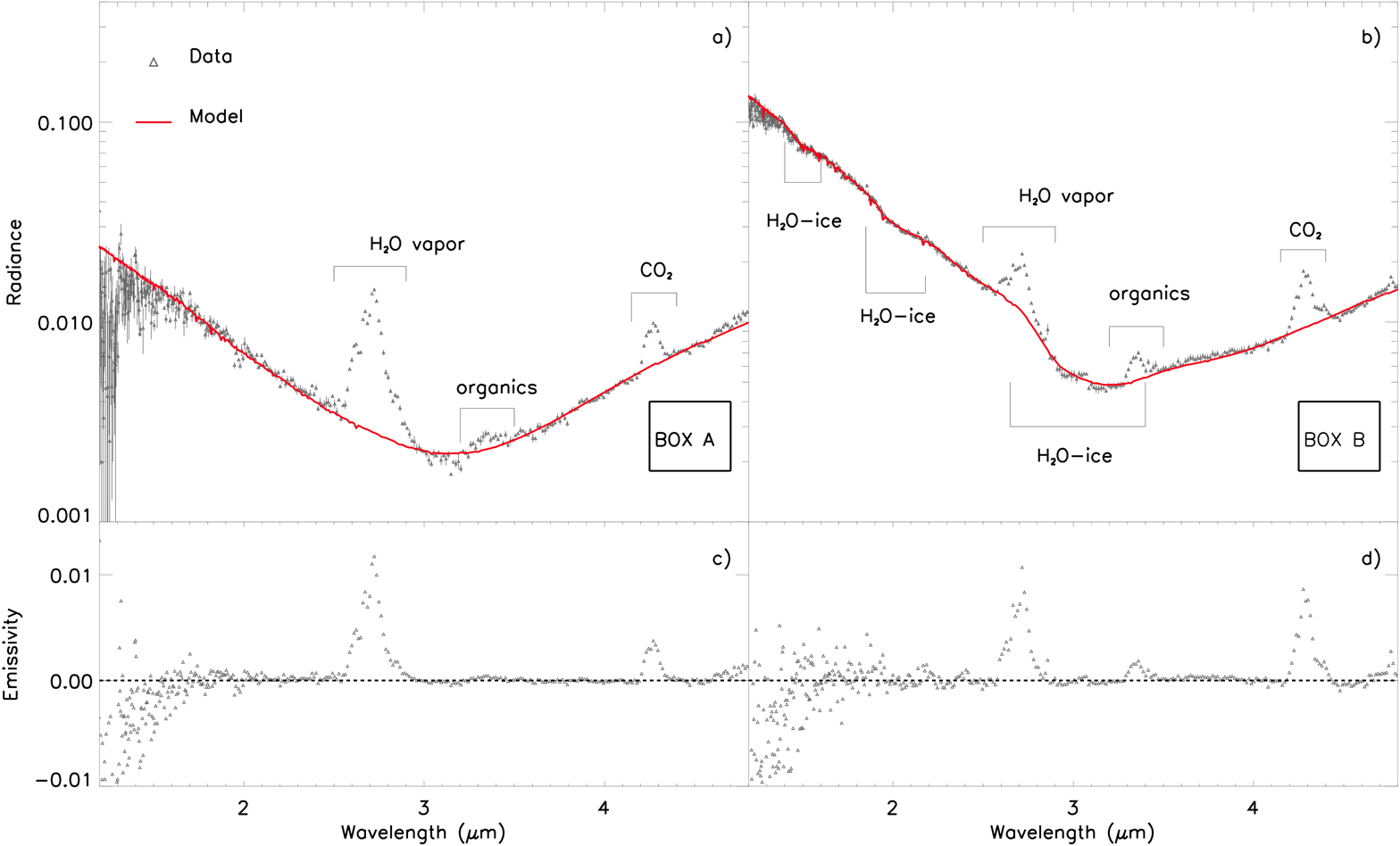}
	\fi
	\caption{Panels (a) and (b) show the radiance spectrum (triangles) of comet 103P/Hartley 2 over the wavelength range 1.2 -- 4.8 $\mu$m, extracted in boxes A and B in panel (a) of Figure \ref{errorRadiance}, respectively. Overplotted is the model (solid red line) shown in Figure \ref{continuum_modeling} converted into radiance. Both spectra clearly display the water vapor, organic, and carbon dioxide emissions, which are not fit by the model, as expected. Panels (c) and (d) display the emissivity spectrum or equivalently the residuals between the model and the data shown in panels (a) and (b), respectively. \textit{See the electronic version of the Journal for a color version
of this figure.}}
	\label{continuum_modeling_radiance}%
\end{figure}
\begin{figure}
	\ifincludegraphics
 	\centering
	\includegraphics[width=\textwidth]{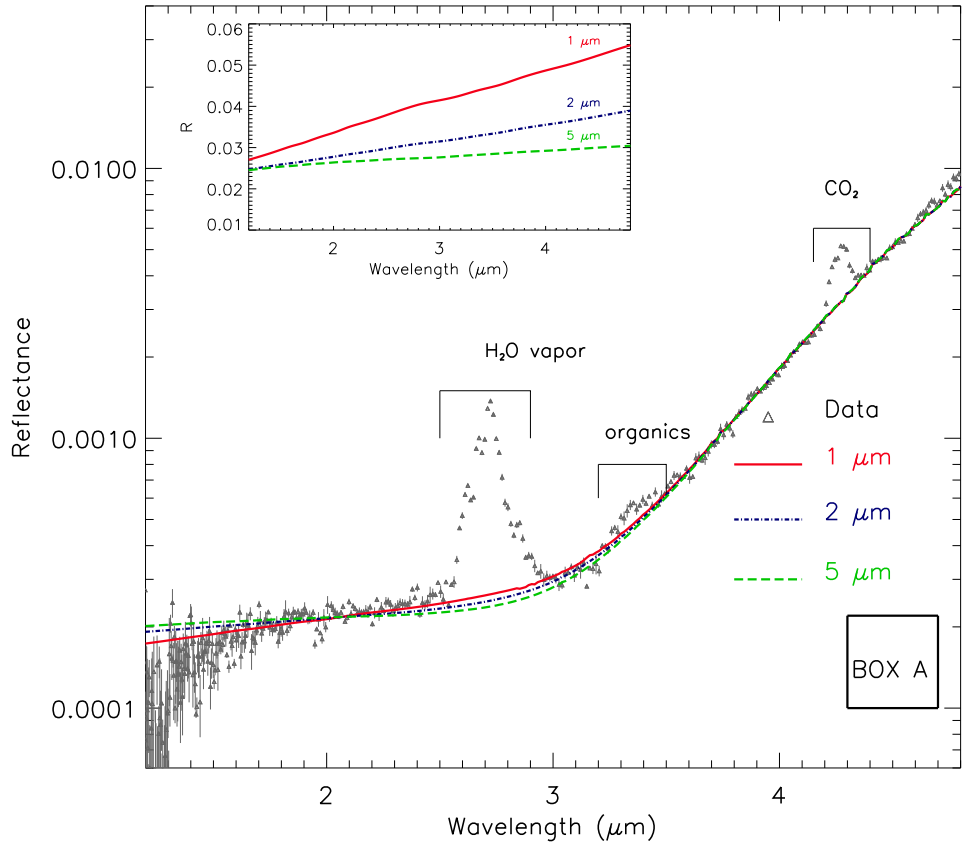}
	\fi
	\caption{The reflectance spectrum of comet 103P/Hartley 2 from the top panel of Figure \ref{continuum_modeling} compared with synthetic spectra obtained assuming amorphous carbon on the order of 1 (solid red line), 2 (dash-dot purple line), and 5 (dashed green line) $\mu$m in diameter. The modeling details are reported in Table \ref{tabB}. The inset panel shows synthetic reflectance spectra
of various amorphous-carbon grains in the
wavelength range covered by the HRI-IR
instrument. \textit{See the electronic version of the Journal for a color version
of this figure.}}
	\label{GrainSizesrefr}%
\end{figure}

\begin{figure}
	\ifincludegraphics
 	\centering
	\includegraphics[width=\textwidth]{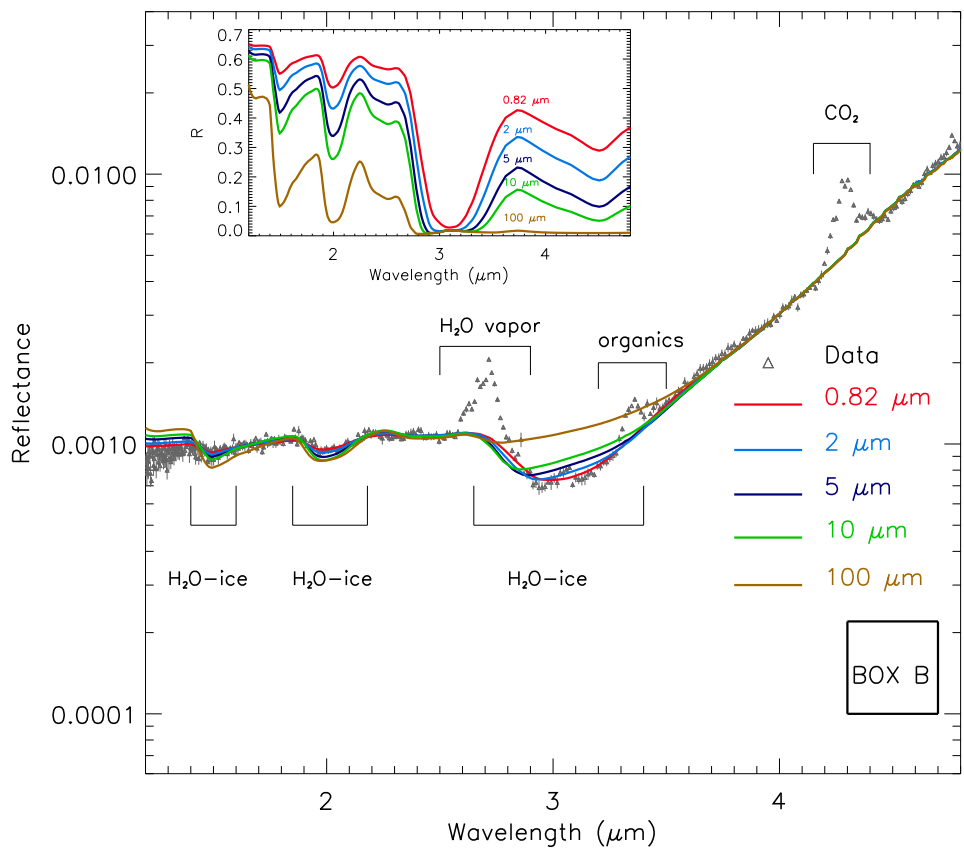}
	\fi
	\caption{The reflectance spectrum of comet 103P/Hartley 2 from the bottom panel of Figure \ref{continuum_modeling} compared with synthetic spectra obtained assuming an areal mixture of amorphous-carbon and water ice grains on the order of 0.82 (red), 2 (blue), 5 (purple), 10 (green), and 100 (brown) $\mu$m. The modeling details are reported in Table \ref{tabB}. The inset panel shows synthetic reflectance spectra
of various sized water ice grains in the
wavelength range covered by the HRI-IR
instrument.}
	\label{GrainSizes}%
\end{figure}
\begin{figure}
	\ifincludegraphics
 	\centering
	\includegraphics[width=\textwidth]{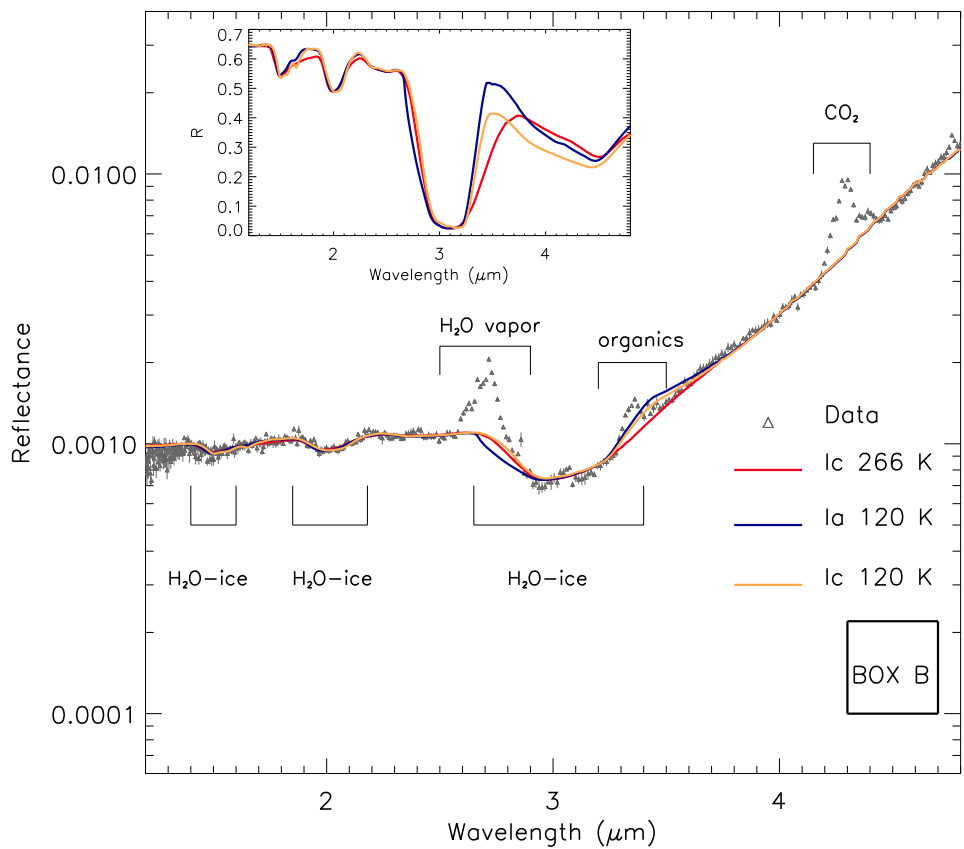}
	\fi
	\caption{The reflectance spectrum of comet 103P/Hartley 2 from the bottom panel of Figure \ref{continuum_modeling} compared with synthetic spectra obtained assuming an areal mixture of amorphous carbon and water ice at different phase and temperatures. We compare models obtained with optical constants from  \citet{Mastrapa2009} of amorphous (Ia) and crystalline (Ic) water ice at 120 K (purple and yellow line, respectively), and with optical constants of \citet{Warren2008} for crystalline ice at 266K (red line). The inset panel shows synthetic reflectance spectra of 100\% water ice (1 $\mu$m particle size) computed using the same set of optical constants.}
	\label{temperature_effects}%
\end{figure}
\begin{figure}
	\ifincludegraphics
 	\centering
	\includegraphics[width=\textwidth]{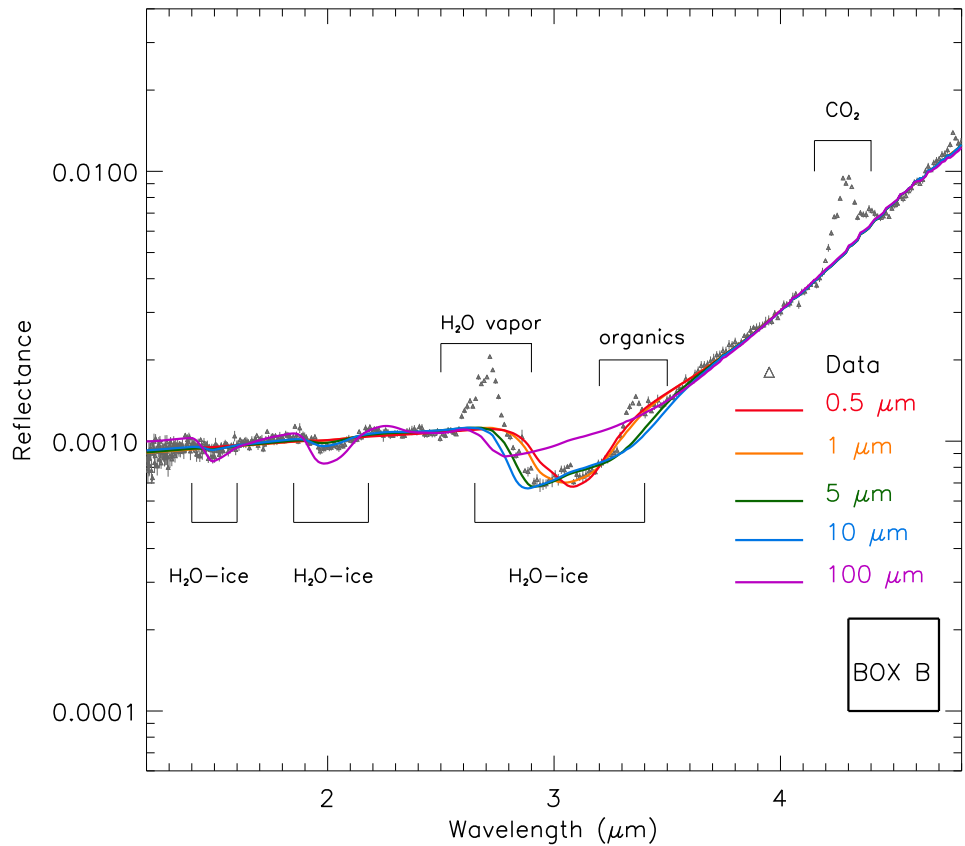}
	\fi
	\caption{Models of the water ice-rich spectrum of Hartley 2 observed 7 min post-CA assuming an intimate mixture. All sizes of water ice lead to poor solutions
when intimately mixed with dark refractory materials.}
	\label{areal_intimate}%
\end{figure}
\begin{figure}
	\ifincludegraphics
 	\centering
	\includegraphics[width=10cm]{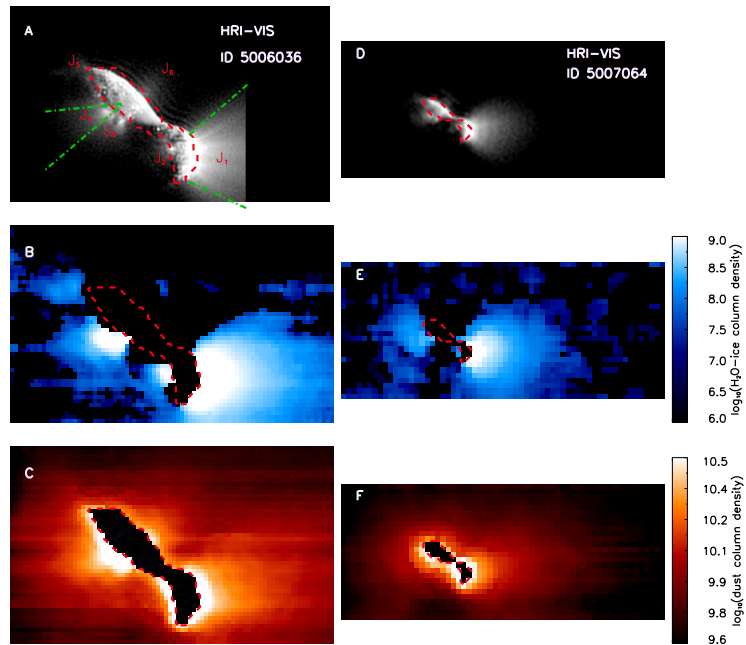}
	\fi
	\caption{Two sets of observations of Hartley 2 separated by 16 min. Water ice and refractories are detected in the innermost coma of Hartley 2. The Sun is to the right. The dashed red line contours the masked comet nucleus.  (A to C) Data acquired 7 min post-CA. (A) HRI-VIS context image at $\sim$11 m/pixel. Overlaid in dash-dot green are the jet areas sampled to produce the profiles in Figure \ref{profiles}. (B) Water ice column density  expressed in number of particles per square meter derived from the 5006000 HRI-IR scan at 55 m/pixel. The lower limit for water ice detection is 6$\times$10$^{6}$ particles/m$^{2}$. The regions in black contain little to no water ice (see text for details). (C) Dust column density expressed in number of particles per square meter. (D to F) Data acquired 23 min post-CA; images matching (A) to (C) at a spatial resolution of 37 m/pixel in the VIS (ID 5007064) and 173 m/pixel in the IR (scan 5007002). \textit{See the electronic version of the Journal for a color version
of this figure.}}
	\label{column density}%
\end{figure}

\begin{figure}
	\ifincludegraphics
 	\centering
	\includegraphics[width= 11cm]{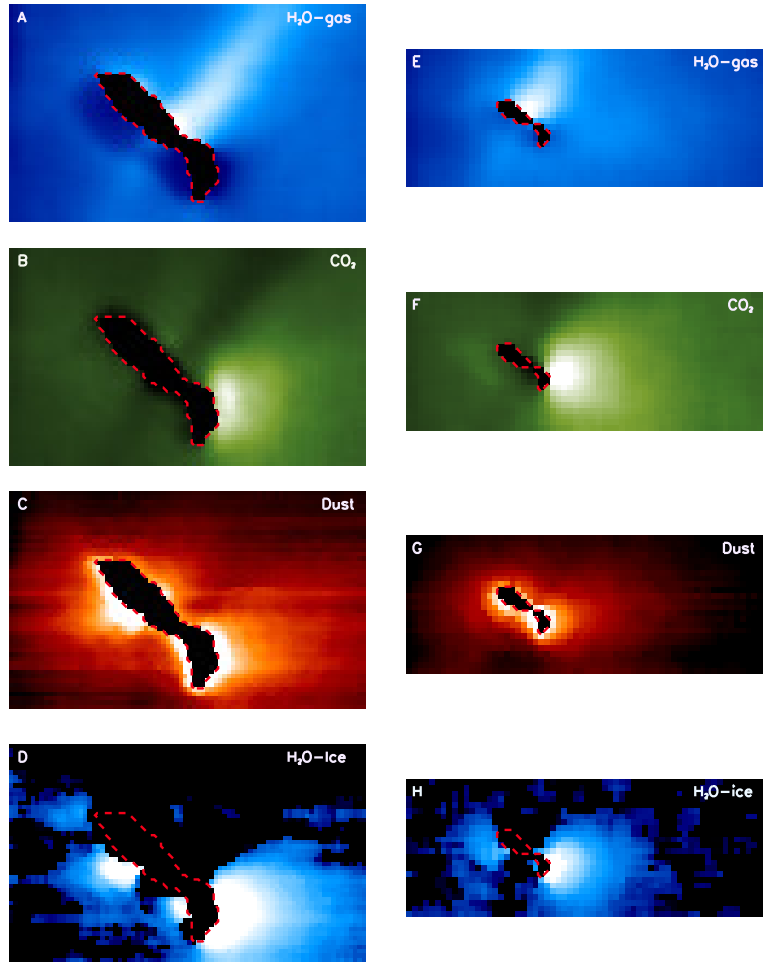}
	\fi
	\caption{Distribution of H$_{2}$O-gas, CO$_{2}$, Dust, and H$_{2}$O-ice in the innermost coma of Hartley 2 as observed 7 min (left column) and 23 min (right column) post-CA. The Sun is to the right. The dashed red line contours the masked comet nucleus. Enhancements in the CO$_{2}$ and H$_{2}$O-gas relative abundances are not correlated. On
the other hand, a strong correlation between CO$_{2}$, H$_{2}$O-ice, and dust is observed
in the inner coma. In all panels
white is relatively more abundant. \textit{See the electronic version of the Journal for a color version
of this figure.}}
	\label{activity}%
\end{figure}

\begin{figure}
	\ifincludegraphics
	\centering
	\includegraphics[width= \textwidth]{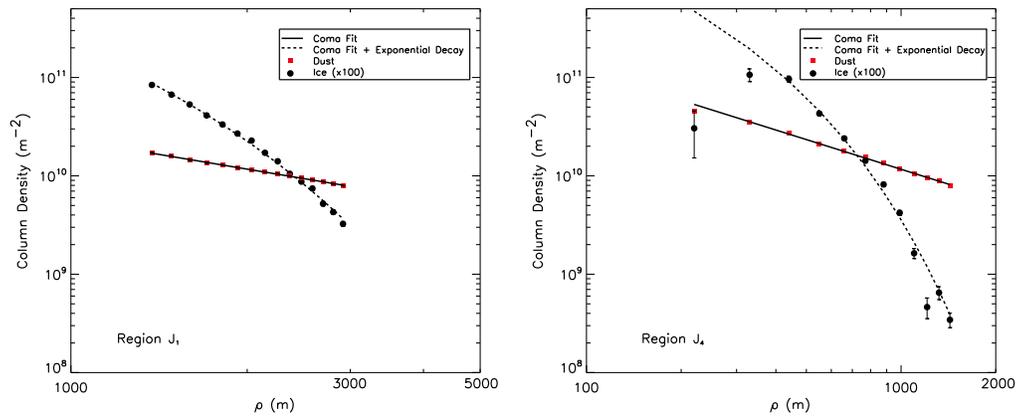}
	\fi
	\caption{Column density profiles of the dust (squares) and water ice (circles) in the jets J$_{1}$ (left panel) and J$_{4}$ (right panel) delimited by the dash-dot green lines shown in panel (A) of Figure \ref{column density}. Solid and dashed lines represent the models of the dust and water ice profiles, respectively. While the dust presents a constant outflow profile, a steeper profile is displayed by the water ice, which can be fit by an exponential decay (see text for details). \textit{See the electronic version of the Journal for a color version
of this figure.}}
	\label{profiles}
\end{figure}











\end{document}